\documentclass[twocolumn]{aastex}

\begin{document}

\title{A Statistical Analysis of Magnetic Field Changes in the Photosphere during Solar Flares Using High-cadence Vector Magnetograms and Their Association with Flare Ribbons}

\author{Rahul Yadav}
\affiliation{Laboratory for Atmospheric and Space Physics, University of Colorado, Boulder, CO 80303, USA; \textnormal{rahul.yadav@lasp.colorado.edu}}

\author{M. D. Kazachenko}
\affiliation{Laboratory for Atmospheric and Space Physics, University of Colorado, Boulder, CO 80303, USA; \textnormal{rahul.yadav@lasp.colorado.edu}}
\affiliation{ National Solar Observatory, 3665 Discovery Drive, 80303, Boulder, CO, USA}
\affiliation{Dept. of Astrophysical and Planetary Sciences, University of Colorado Boulder, 2000 Colorado Ave, 80305, Boulder, CO, USA}

\begin{abstract}
We analyze high-cadence vector magnetograms (135~s) and flare-ribbon observations of 37 flares from the Solar Dynamics Observatory to understand the spatial and temporal properties of changes in the photospheric vector magnetic field and their relationship to footpoints of reconnected fields. Confirming previous studies, we find that the largest permanent changes in the horizontal field component lie near the polarity inversion line, whereas changes in the vertical field are less pronounced and are distributed throughout the active region. We find that pixels swept up by ribbons do not always exhibit permanent changes in the field. However, when they do, ribbon emission typically occurs several minutes before the start of field changes. The changes in the properties of the field show no relation to the size of active regions, but are strongly related to the flare-ribbon properties such as ribbon magnetic flux and ribbon area. For the first time, we find that the duration of permanent changes in the field is strongly coupled with the duration of the flare, lasting on average 29\% of the duration of the GOES flare. Our results suggest that changes in photospheric magnetic fields are caused by a combination of two scenarios: contraction of flare loops driven by magnetic reconnection and coronal implosion.
 \end{abstract}

\keywords{ Sun: Magnetic fields -- Sun: flares}

\section{Introduction}
Solar flares are one of the most spectacular and energetic phenomena on the Sun. They result in the intense release of radiation across the electromagnetic spectrum, affecting different layers of the solar atmosphere. Frequently, strong flares are accompanied by coronal mass ejection (CME), releasing a large amount of radiation that may have severe space-weather impacts \citep{2015SpWea..13..524S}. Therefore, understanding the real mechanism behind explosive events like solar flares and CMEs has become one of the hot topics in solar physics research \citep{2008LRSP....5....1B, 2022SoPh..297...59K}.

According to the standard flare model, solar flares are caused by the magnetic reconnection or reconfiguration of field lines in the corona. During the reconnection processes the magnetic energy is converted into primarily kinetic and thermal energies driving the acceleration of the particles into the lower solar atmosphere \citep{1964NASSP..50..451C, 1966Natur.211..695S, 1974SoPh...34..323H, 1976SoPh...50...85K}. The deposition of energy gives rise to intense brightening and emission of hard and soft X-rays in the lower solar atmosphere. The appearance of bright structures in the chromosphere or transition region is normally referred to as flare ribbons. These ribbons indicate footpoints of reconnected field lines. Their morphology is frequently utilized to trace the evolution of coronal magnetic energy release \citep{Longcope2007,Kazachenko2012, 2017ApJ...838...17Q}.
As the coronal magnetic field lines are rooted in the photosphere, the investigation of magnetic field topology and the evolution of field lines from the lower solar atmosphere up to the corona is required to understand different aspects of a flare \citep{2011SSRv..158....5H}.

Flare observations have demonstrated that an intense flare can distort the structure of active regions (ARs), rapidly rotate sunspots in the photosphere \citep{2014ApJ...782L..31W,2016NatCo...713104L, 2018NatCo...9...46X}, and can lead to contraction and oscillation of coronal loops \citep{ 2015A&A...581A...8R, 2016ApJ...833..221W}. In the past, many efforts have been made to understand the flare-related changes in the photospheric magnetic field  (e.g. \citealt{1964NASSP..50...95S, 1970IzKry..41...97Z, 1984ApJ...276..379M, 1993ApJ...407L..89W, 1994ApJ...424..436W, 1999SoPh..190..459K, 1999ApJ...525L..61C, 2002ApJ...572.1072S, 2002ApJ...576..497W, 2004ApJ...605..546Y}).
During the last decade, the availability of high-cadence photospheric vector magnetograms from  ground-based and space-based telescopes, such as the \textit{Solar Dynamics Observatory} (SDO; \citealt{2012SoPh..275....3P}), have provided us evidence of rapid and permanent changes in the longitudinal and transverse magnetic fields associated with solar flares in the photosphere \citep{2005ApJ...635..647S, 2010ApJ...724.1218P,2012ApJ...745L..17W,2012ApJ...749...85G,2017ApJ...839...67S, 2018ApJ...852...25C, 2019ApJS..240...11P, 2022ApJ...934L..33L}. Recently, magnetic field changes in the chromosphere, in addition to the photosphere, have also been reported from flare observations performed at ground-based observatories \citep{2017ApJ...834...26K, 2021A&A...649A.106Y}. 

Observational evidence of an increase in the horizontal component of the magnetic field in the photosphere and the contraction of coronal loops during a flare is generally interpreted with the conjecture proposed by \cite{2000ApJ...531L..75H}, also known as \textit{coronal implosion}. It states that during a transient event, such as a flare or a CME, in a low plasma-$\beta$ atmosphere with negligible gravity, the coronal field lines must contract in such a way as to reduce the magnetic energy, $E_{mag} = \int_{V}^{} B^2/8\pi dV$.
The release of the free magnetic energy should be accompanied by a decrease in the magnetic pressure and volume, which can lead to loop contraction at the flare sites \citep{2008ASPC..383..221H, 2012SoPh..277...59F}. The loop contraction during flares has been noticed in numerous observations (e.g. \citealt{2009ApJ...696..121L, 2013ApJ...777..152S}). Such coronal magnetic implosion or loop contractions could increase the horizontal component of the magnetic field in the photosphere near the polarity inversion line (PIL). 

Numerical studies have also been performed by various authors to understand the mechanism behind the loop contractions and related changes in the field. \citealt{2011ApJ...727L..19L} analyzed flare-associated magnetic field changes in observations and simulations. They found that both observations and simulations show an increase in the horizontal component of the magnetic field
near the PIL after the flare. They argued that these changes are the result of the collapse of the preexisting coronal flux rope and the subsequent implosion of the magnetic field lines above the PIL, consistent with the prediction by \citealt{2008ASPC..383..221H}.  
In a 3D magnetohydrodynamic model of
an erupting magnetic-flux rope, \citealt{2017ApJ...837..115Z} found that vortices developed on both sides of the expanding flux-rope footpoints during a flare eruption could  cause the loop contraction. Within the framework of ideal magnetohydrodynamics, \citet{2017ApJ...851..120S} found that the dynamics of loop implosion are also sensitive to the velocity disturbance generated close to the reconnection site. 
Recently, \cite{2019ApJ...877...67B} performed a generic 3D magnetohydrodynamics simulation of an eruptive flare to understand the mechanism behind the changes in the field. They found that enhancements in the photospheric horizontal magnetic fields are due to the contraction of sheared flare loops caused by magnetic reconnection, which contradicts previous interpretations based on the implosion conjecture.

During the last decades,  the photospheric magnetograms available from various space-based instruments have improved our understanding of the photospheric changes associated with flares. However, most of the previous studies were performed with a low cadence or focused more on the longitudinal changes. For example, vector magnetograms obtained from the Helioseismic and Magnetic Imager (HMI)/SDO \citep{2012SoPh..275..207S} have a cadence of 12 minutes, which is not sufficient to temporarily resolve the fast changes that normally occur during a flare.

In this study, we present a statistical analysis of flares to understand their magnetic imprints in the photosphere using high-cadence (135~s) vector magnetograms
obtained from HMI/SDO. We aim to clarify how the characteristics of changes in the photospheric field are related to ultraviolet (UV) emissions and ribbon morphology, which are the footpoints of reconnected field lines.

The remainder of this paper is structured as follows. In Section~\ref{sec: data and method}, we describe our data set and methods employed to characterize the changes in the field. We present our results in Section~\ref{sec_results}. We then discuss them in Section~\ref{sec: Discussion}. Finally, in Section~\ref{sec: conclusion}, we summarize our conclusions.

\section{Data and Methods}
\label{sec: data and method}
\begin{figure}[!t]
    \centering
    \includegraphics[width=1\linewidth]{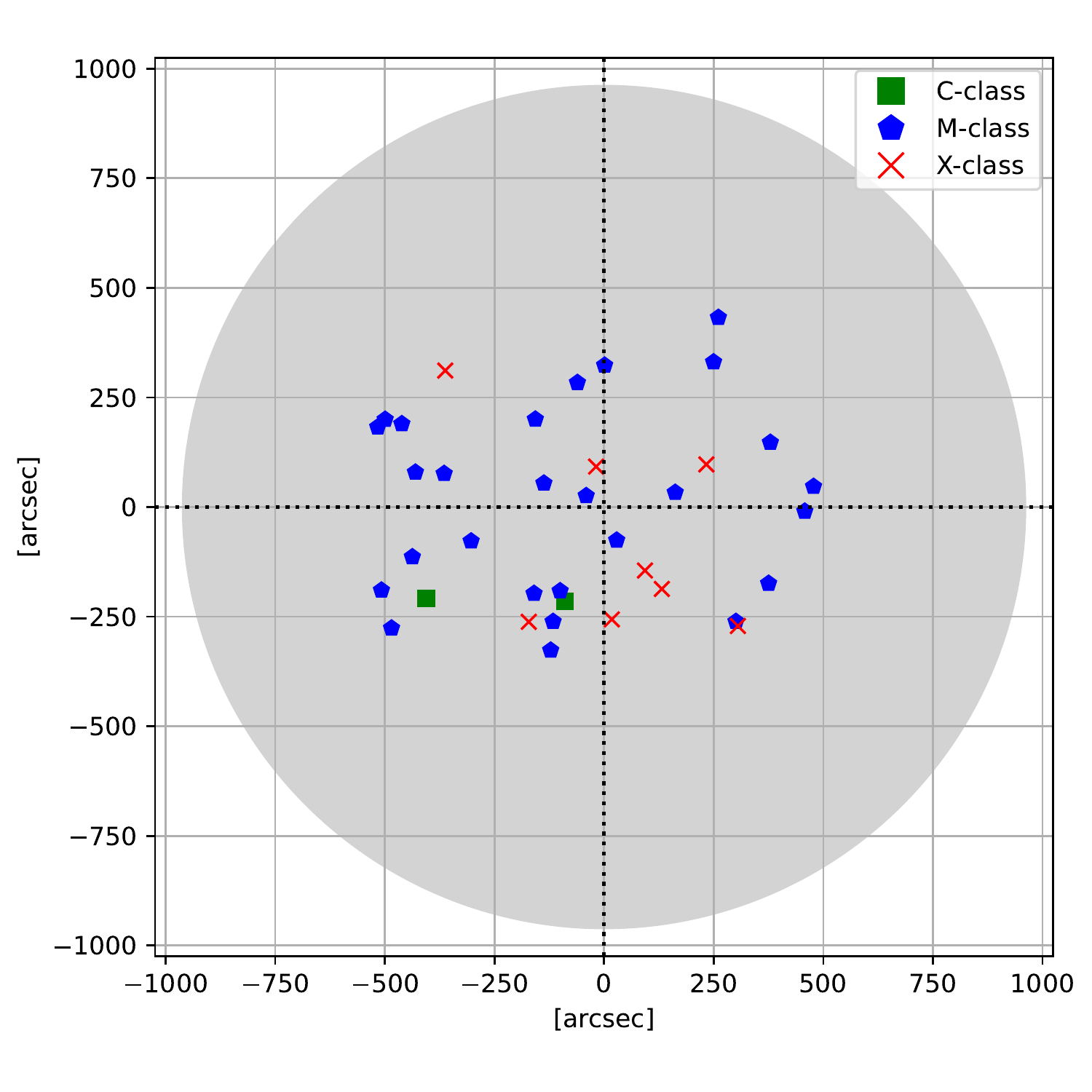}
    \caption{Location of selected flare events on the artificial solar disk. Green square, blue pentagon, and red cross symbols indicate the position of C-, M-, and X-class solar flares on the solar disk, respectively.}
    \label{flareloc}
\end{figure}
\label{sec-observation}

\begin{table*}[]
    \centering
        \caption{List of 37 flares. For each event we show the following flare properties: NOAA AR number, location on the solar disk, total cumulative ribbon area (S$_\mathrm{rbn}$), total AR area (S$_{AR}$), total unsigned magnetic flux in ribbons ($\Phi_{rbn}$) and in the AR ($\Phi_{AR}$), duration of the GOES flare ($\tau_{GOES}$), and ribbon distance (d$_\mathrm{rbn}$). The area is expressed in millionths of the solar hemisphere, which is equivalent to 3$\times$10$^6$ km$^2$.} 
    \begin{tabular}{lcccccccccccc}
        \hline
    \hline
Event &	Flare start	time & NOAA	& Flare  &  Location  & S$_\mathrm{rbn}$ & S$_\mathrm{AR}$  &$|\Phi_\mathrm{rbn}|$ &$|\Phi_\mathrm{AR}|$  & $\tau_{GOES}$ & d$_\mathrm{rbn}$ \\
no. &       (UT)         & AR no.& Class  & on Disk  &  (MSH) & (MSH) & {(10$^{21}$ Mx}) & {(10$^{21}$ Mx}) & (min.) & (Mm)  \\
\hline
1 & 2010-08-07T17:55 & 11093 & M1.0 & N12E31 & 317 & 2857 & 4.8 & 30.3 & 51.2 & 62.5 \\
2 & 2011-02-15T01:43 & 11158 & X2.2 & S20W10 & 512 & 1863 & 11.6 & 30.7 & 22 & 27.1 \\
3 & 2011-08-03T13:17 & 11261 & M6.0 & N16W30 & 370 & 3495 & 7.6 & 43.4 & 52.4 & 51.0 \\
4 & 2011-09-06T22:11 & 11283 & X2.1 & N14W18 & 436 & 2436 & 5.9 & 26.4 & 12 & 9.6 \\
5 & 2011-10-02T00:37 & 11305 & M3.9 & N12W26 & 178 & 1387 & 2.4 & 16.7 & 21.2 & 15.6 \\
6 & 2011-11-15T12:29 & 11346 & M1.9 & S18E26 & 215 & 8098 & 3.5 & 52.9 & 19.6 & 14.4 \\
7 & 2011-12-27T04:11 & 11386 & C8.9 & S17E23 & 139 & 4366 & 2.0 & 40.6 & 19.2 & 26.1 \\
8 & 2012-01-23T03:37 & 11402 & M8.7 & N33W21 & 892 & 8002 & 17.0 & 94.5 & 55.6 & 51.8 \\
9 & 2012-03-07T00:01 & 11429 & X5.4 & N18E31 & 1173 & 6152 & 30.4 & 77.1 & 37.6 & 51.3 \\
10 & 2012-03-09T03:21 & 11429 & M6.3 & N15W03 & 768 & 3920 & 14.4 & 57.1 & 55.6 & 30.8 \\
11 & 2012-03-10T17:15 & 11429 & M8.4 & N17W24 & 904 & 4412 & 16.9 & 61.9 & 74.4 & 43.6 \\
12 & 2012-03-14T15:07 & 11432 & M2.8 & N14E01 & 234 & 1933 & 3.1 & 18.6 & 27.6 & 18.8 \\
13 & 2012-07-12T15:37 & 11520 & X1.4 & S13W03 & 428 & 6335 & 8.6 & 85.9 & 113 & 53.0 \\
14 & 2012-11-21T06:45 & 11618 & M1.4 & N08W00 & 194 & 2104 & 3.4 & 26.0 & 22.4 & 27.6 \\
15 & 2013-04-11T06:55 & 11719 & M6.5 & N07E13 & 324 & 2959 & 4.5 & 29.2 & 33.2 & 17.4 \\
16 & 2013-05-16T21:35 & 11748 & M1.3 & N11E37 & 222 & 5779 & 3.8 & 43.4 & 26.8 & 33.5 \\
17 & 2013-05-31T19:51 & 11760 & M1.0 & N12E37 & 151 & 6568 & 2.1 & 36.0 & 13.6 & 9.7 \\
18 & 2013-08-17T18:49 & 11818 & M1.4 & S07W32 & 313 & 4395 & 6.1 & 50.2 & 65 & 33.4 \\
19 & 2013-12-28T17:53 & 11936 & C9.3 & S16E07 & 105 & 2330 & 1.4 & 24.5 & 14.4 & 11.4 \\
20 & 2014-01-07T18:03 & 11944 & X1.2 & S12W08 & 792 & 5075 & 11.6 & 73.4 & 53.6 & 102.0 \\
21 & 2014-01-31T15:31 & 11968 & M1.1 & N09E29 & 317 & 7638 & 3.4 & 63.2 & 20.8 & 51.0 \\
22 & 2014-02-01T07:13 & 11967 & M3.0 & S14E17 & 222 & 5993 & 5.5 & 92.7 & 21.6 & 65.9 \\
23 & 2014-02-12T03:51 & 11974 & M3.7 & S12W11 & 408 & 3052 & 6.7 & 37.8 & 45.6 & 63.3 \\
24 & 2014-03-20T03:41 & 12010 & M1.7 & S15E27 & 228 & 5029 & 3.4 & 49.9 & 25.6 & 46.3 \\
25 & 2014-08-01T17:55 & 12127 & M1.5 & S09E08 & 409 & 3809 & 5.2 & 38.6 & 52.4 & 46.5 \\
26 & 2014-08-25T14:45 & 12146 & M2.0 & N06W39 & 221 & 3870 & 4.5 & 42.3 & 44.8 & 14.4 \\
27 & 2014-08-25T20:05 & 12146 & M3.9 & N07W43 & 253 & 4494 & 6.0 & 45.4 & 22.8 & 14.8 \\
28 & 2014-09-08T23:11 & 12158 & M4.5 & N16E26 & 309 & 3166 & 8.4 & 42.6 & 138 & 36.0 \\
29 & 2014-09-10T17:21 & 12158 & X1.6 & N11E05 & 702 & 2374 & 12.2 & 30.0 & 58.4 & 38.7 \\
30 & 2014-09-28T02:39 & 12173 & M5.1 & S13W23 & 361 & 5890 & 7.0 & 79.9 & 39.2 & 53.2 \\
31 & 2014-10-22T14:01 & 12192 & X1.6 & S14E13 & 811 & 9632 & 17.7 & 155.9 & 47.6 & 70.9 \\
32 & 2014-12-17T00:57 & 12242 & M1.5 & S20E08 & 151 & 5538 & 2.8 & 67.8 & 22.4 & 50.0 \\
33 & 2014-12-17T04:25 & 12242 & M8.7 & S18E08 & 379 & 5639 & 8.5 & 69.7 & 54.4 & 32.0 \\
34 & 2014-12-18T21:41 & 12241 & M6.9 & S11E10 & 461 & 3622 & 9.3 & 48.6 & 43.2 & 15.6 \\
35 & 2014-12-20T00:11 & 12242 & X1.8 & S19W29 & 1289 & 7974 & 26.5 & 113.6 & 43.2 & 53.6 \\
36 & 2015-11-04T13:31 & 12443 & M3.7 & N06W10 & 535 & 3364 & 6.9 & 39.0 & 41.2 & 45.7 \\
37 & 2015-11-09T12:49 & 12449 & M3.9 & S12E33 & 476 & 7828 & 8.4 & 55.3 & 38.4 & 25.2 \\
\hline
    \end{tabular}

    \label{tab_flareinfo}
\end{table*}

In this study, we selected 37 flaring events, including 8 X-, 27 M-, and 2 C-class flares in 31 ARs listed in Table~\ref{tab_flareinfo}. Out of 37 flares, 31 were eruptive whereas six flares were confined or non-eruptive (event nos: 7, 16, 19, 22, 31, and 32 in Table~\ref{tab_flareinfo}). These events are taken from the FlareMagDB\footnote{\url{http://solarmuri.ssl.berkeley.edu/~kazachenko/FlareMagDB/}} catalog created by \cite{2022ApJ...926...56K}. As shown in the Figure \ref{flareloc}, the selected events are distributed within 45$^\circ$ from the disk center and occurred from 2010 August to 2015 November (see Table~\ref{tab_flareinfo}).
For each event, we used the high-cadence (135 s) full-disk vector magnetograms obtained from the HMI \citep{2012SoPh..275..207S} on board SDO \citep{2012SoPh..275....3P}. 
The HMI samples the spectral region around the \ion{Fe}{1}~6173.3~\AA~absorption line at six wavelength points with a bandwidth of 76~m\AA~and records a full set of Stokes parameters (I, Q, U, V) in 135~s with a pixel size of 0\arcsec.5. The post-processing and data acquisition of 135~s cadence vector magnetograms are described in \cite{2017ApJ...839...67S}. 
The full-disk vector magnetogram\footnote{\url{http://jsoc.stanford.edu/ajax/lookdata.html?ds=hmi.B_135s}} is retrieved by inverting a full set of Stokes parameters using the Milne-Eddington inversion approach \citep{2011SoPh..273..267B}. To resolve the azimuthal ambiguity we employed the {\tt hmi\_disambig.pro} routine of the HMI SolarSoft package. After this 180-ambiguity correction, we transformed the magnetic field vector inferred in the line-of-sight frame to the solar local reference frame using the transformation matrix given by \cite{1990SoPh..126...21G}.

\cite{2017ApJ...839...67S} have carried out a comparison between pairs of 135 and 720~s full-disk vector magnetograms retrieved from HMI. They found that the 135 and 720 s data agree well in the strong-field regions (B~$>$~300~G). However, in comparison to 720~s data, the 135~s data have higher noise due to the shorter integration time. 

For each flaring event, in addition to HMI vector magnetograms, we also used a sequence of 1600~\AA\ images, obtained from the Atmospheric Imaging Assembly (AIA; \citealt{Lemen2012}) on board SDO, at the cadence of 24~s with a pixel size of 0\arcsec.6~covering the full evolution of the flare ribbons. These AIA~1600~\AA\ images are used to trace the morphology of flare ribbons. Then we used the {\tt aia\_prep.pro} routine of the SolarSoft package to align the AIA image sequences with the HMI vector magnetograms. For further analysis, we defined the region of interest centered on the AR and covering the entire flare-ribbon area. The final co-aligned AIA and HMI data consists of a cube of a 480\arcsec~$\times$~480\arcsec~field of view (FOV) with a 0\arcsec.6~pixel scale.

To characterize and study the evolution of the changes in the field, we selected a $2$~hr interval around the time of the flaring, 1~hr before and after the GOES flare peak time. The flare start, peak, and end times are identified using \textit{Geostationary Operational Environmental Satellite} (GOES) 1 – 8~\AA\ X-ray flux. 
The magnetic field map sequence of $2$~hr with a cadence of $135$~s is sufficient to capture the permanent changes in the field in all events, except the cases, where the flaring duration was above $75$ minutes (event nos. 11, 13, and 28 in Table \ref{tab_flareinfo}). For this event we selected a $3$~hr interval (1.5~hr on either side of the flare peak time). 

\subsection{Desaturation of AIA 1600~\AA\ images}
We used SDO/AIA 1600~\AA\ images and HMI magnetograms to compare UV emissions from the chromosphere to the magnetic field changes in the photosphere during 37 selected flares. The energy released during flares in the chromosphere and the transition region gives rise to emissions in the flare ribbons. If the heating caused by a flare is sufficiently strong, then the AIA~1600~\AA\ pixels located on the flare ribbon get saturated due to the diffraction patterns from the EUV-telescope entrance filter and CCD saturation. 

To correct the saturated intensities of the pixels we employed the method given by \cite{2017ApJ...845...49K}. With this approach, we first identify the pixels above a threshold intensity of (5000~counts~s$^{-1}$) and the neighboring 2 and 10 pixels in the $x$- and $y$-directions, respectively. In the next step, we replace all selected pixel intensities with the value obtained from linear interpolation in time between the previous and the following image sequences when the pixels are unsaturated. More details regarding this method are given in \cite{2017ApJ...845...49K}. For further analysis, we used the saturation-corrected images. 

\subsection{Determination of Permanent Changes in the Magnetic Field}
\label{sec_stepfit}

\begin{figure}
    \centering
    \includegraphics[width=1\linewidth]{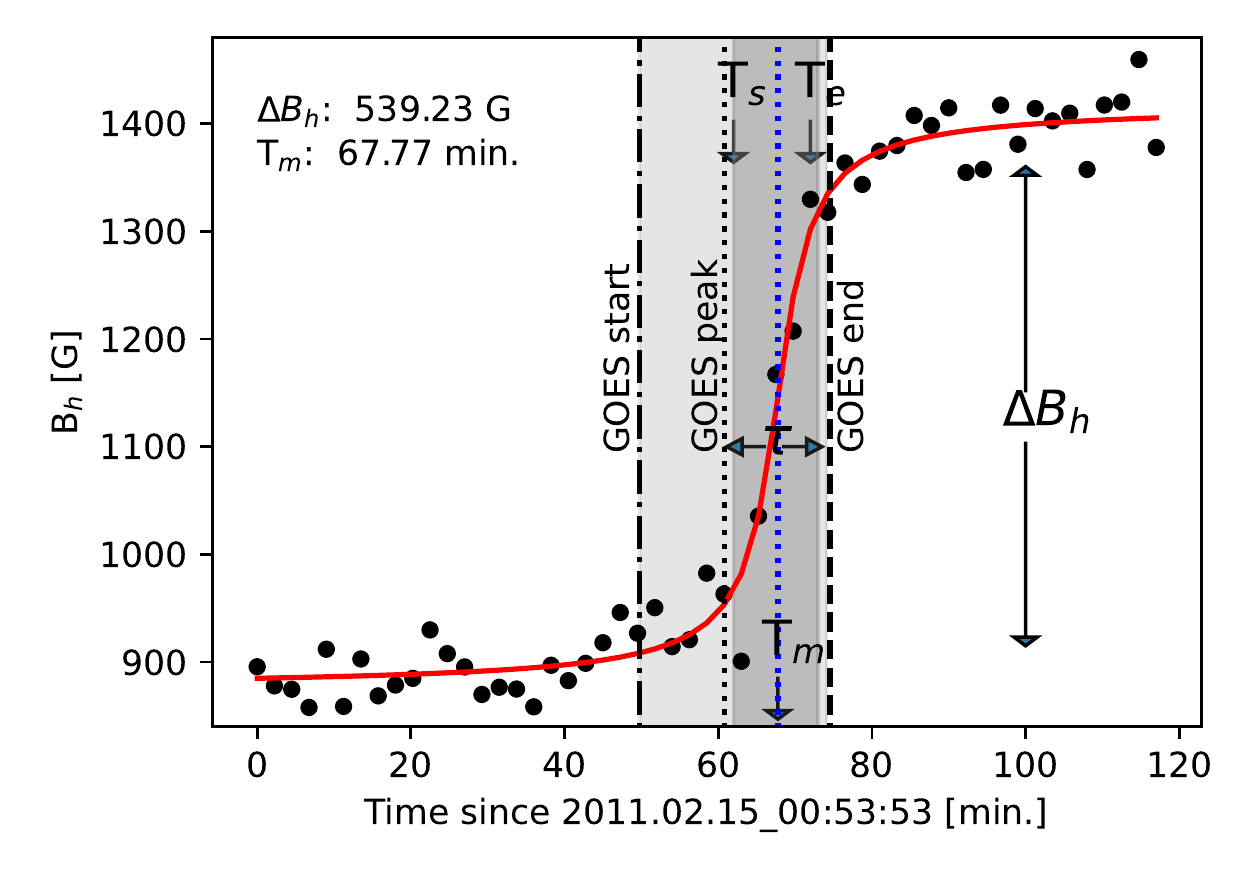}
    \caption{Illustration of the parameters that we used to describe the changes in the magnetic field in a single pixel (see Equation \ref{eq_stepfit}). The temporal evolution of the horizontal magnetic field during an X2.2 flare for a single pixel is indicated by black circles, whereas the red line indicates the best fit of Equation \ref{eq_stepfit}. The start time (T$_s$) and end (T$_e$) time of the change in the field are indicated by black arrows. The dark gray shaded area indicates the duration of the change in the field ($\tau$), whereas the total light and dark gray shaded area refer to the duration of the GOES X-ray flaring. $\Delta B_h$ refers to the measure of the change in the field. The dotted blue line indicates the mid-time of change in the field (T$_m$). Dotted dashed, dotted, and dashed black lines indicate GOES flare start, peak, and end times, respectively.}
    \label{fig:step_fit_demo}
\end{figure}
To determine and characterize the permanent changes in the field in the selected FOV, we fitted the co-aligned time sequences of the horizontal ($B_h = \sqrt{B_x^2 + B_y^2}$) and vertical ($B_z$)  components of the magnetic field in each pixel with a step-like function \citep{2005ApJ...635..647S},
\begin{equation}
    B_i(t) = a + bt +c\Bigl\{1 + \frac{2}{\pi}\textrm{tan}^{-1}[n(t - T_m)]\Bigl\},
\label{eq_stepfit}
\end{equation}

where $a + bt$ describes a linear evolution of the background field with time $t$, $c$ represents the half amplitude of the step, $n$ is the inverse of the time interval controlling the slope of the step, and $T_m$ is the time corresponding to the midpoint of the step. 

The temporal evolution of each pixel is then fitted by varying the free parameters: $a, b, c, n,$ and $T_m$. To fit the step-like function we used the Levenberg-Marquardt method of nonlinear least-squares minimization. 
The best-fitted parameters were then used to characterize the changes in the field and to create maps of fitted parameters: $\Delta B_i$, $\tau$, $\Delta \dot B$, T$_m$, T$_s$, and T$_e$. The $\Delta B_i = 2c$ is a measure of the change in the magnetic field (in Gauss).
The $\tau = \pi n^{-1}$ corresponds to the period of time over which the change in the magnetic field occurs or the duration of the change in the field (in minutes). The $\Delta \dot B=\Delta$B/$\tau$ corresponds to the rate of the change in the magnetic field (in G/min). The T$_m$ is the mid time of the change in the field. The start and end times of the changes in the field are estimated as $T_s = T_m - \tau/2$ and $T_e = T_m + \tau/2$, respectively. These derived parameters are illustrated in Figure \ref{fig:step_fit_demo}.

\subsection{Criteria Used to Characterize the Changes in the Magnetic Field}
\label{sec:criteria field change}
 In our data set, we find that there are different types of field evolution in both $B_h$ and $B_z$, which cannot be entirely described by Equation~\ref{eq_stepfit}.
 While some changes are related to flare, some of them may be related to flux emergence or cancellation that could lead to non-linear background evolution of the magnetic field. 
 In this study, we do not investigate the role of flux emergence or cancellation, but focus only on the permanent changes in the magnetic field in the selected events. 
To identify the correct characteristics and permanent change in the magnetic field, we apply the following criteria to each pixel fitted using the Eq.~\ref{eq_stepfit}: 
\begin{enumerate}
    \item The pixels should reside in the flare-ribbon area or the field strength of  the pixels should be greater than 300~G: $|B|>300~G$ 
    \item The $\Delta B$ value should be greater than 100~G and the maximum $\Delta B$ value should be less than 800~G. We impose these limits to avoid pixels having a value less than the uncertainty of B or having a strong background evolution due to moving magnetic features near the AR \citep{2005ApJ...635..659H}. 
    \item The start and end times of the changes in the field should lie within the duration  of the flaring given by GOES X-ray flux. In some pixels, we find that the changes in the field begins before the flare start time and the change in the field ends after the flare end time. Such pixels ($< 1\%$) are not included in the analysis as they may not be related to flares. 
   Moreover, the mid time of the change in the field, $T_m$, should lie within the GOES flare start and end times. Pixels with a $T_m$ value beyond the duration of the flaring are excluded. 
    \item The change in duration, $\tau$, should be greater than the cadence of HMI vector magnetograms (135~s), even though there are a small number of pixels exhibiting $\tau$ less than 135s.
\end{enumerate}
 The pixels satisfying the above conditions and having the best chi-square values ($< 3$) obtained from Equation~\ref{eq_stepfit} are then used to estimate the following parameters in the selected region: $\Delta B_i$, $\tau$, $\Delta \dot B$, T$_m$, T$_s$, and T$_e$. As an example, the parameters retrieved after fitting a single pixel, using Eq. \ref{eq_stepfit}, are illustrated in Figure \ref{fig:step_fit_demo}. 

Additionally, for each event we also defined the following AR and flare properties (see Table~\ref{tab_flareinfo}): the total AR area (S$_{AR}$), calculated as the area of the pixels having an intensity of less than 85\% of the intensity of the quiet-Sun from limb-darkening corrected continuum images \citep{1997SoPh..175..197P}; the total cumulative ribbon area (S$_\mathrm{rbn}$), the total unsigned magnetic flux in the AR ($\Phi_{AR}$) and the total flux in the cumulative ribbon area ($\Phi_\mathrm{rbn}$) are estimated following the approach given by \cite{2017ApJ...845...49K}; the duration of the GOES flare ($\tau_{GOES}$) is defined as the time difference between the GOES flare start and end times for each event; the ribbon distance (d$_\mathrm{rbn}$) is estimated as the separation between the two magnetic-flux weighted centroids of the ribbons in the positive and negative polarities (see Figure~\ref{fig:appendix_overview_ribbon_sep} in the Appendix).

 \begin{figure}[t!]
    \centering
    \includegraphics[width=1\linewidth]{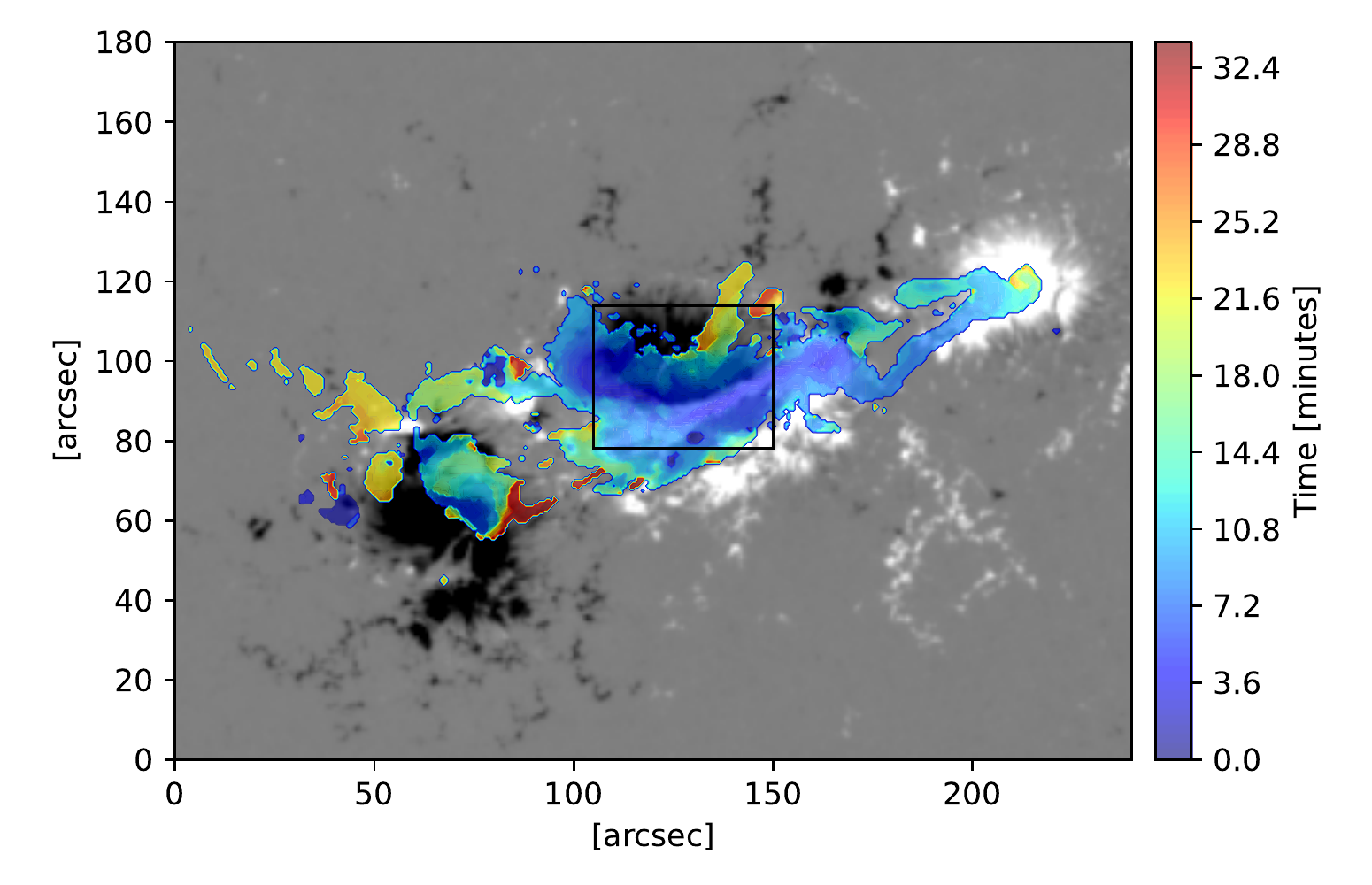}
    \includegraphics[width=0.9\linewidth]{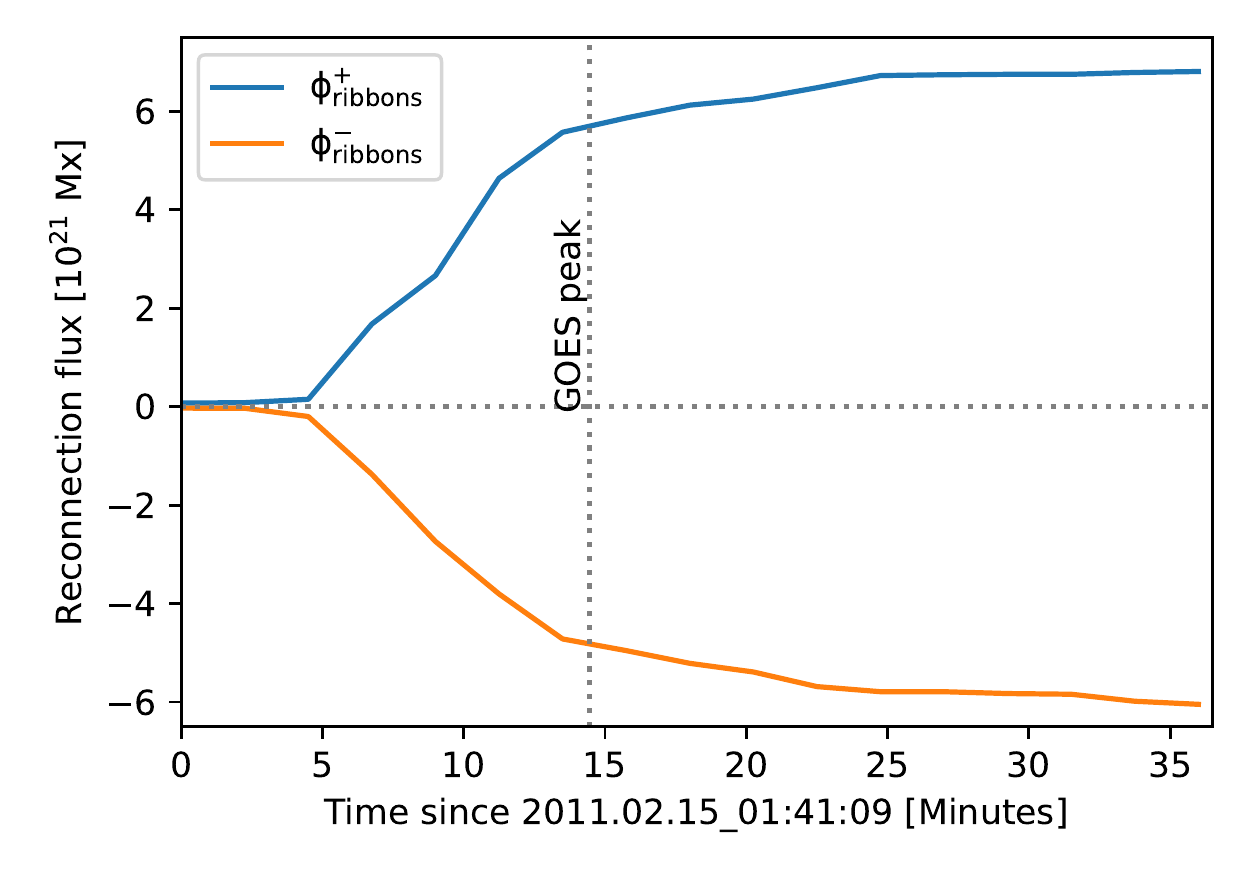}
    \caption{Evolution of flare ribbons during an X2.2 flare on 2011 February 15. \textit{Top panel:} spatial locations and evolution of ribbons color coded in time. The vertical component of the magnetic field ($B_z$) is shown as the background, where black and white colors indicate negative and positive polarities (saturated at $\pm800$~G), respectively. The black box shows the FOV for Figure \ref{fig:fittep_maps}.\textit{ Bottom panel:} time profiles of the total reconnection flux integrated into the positive and negative polarities, respectively. The vertical dotted line indicates the GOES X-ray peak time.}
    \label{fig:ribbon_evolution}
\end{figure}

\section{Results}
\label{sec_results}

\begin{figure*}
    \centering
    \includegraphics[clip,trim=0.cm 0.45cm 0.5cm 0cm, width=0.75\linewidth]{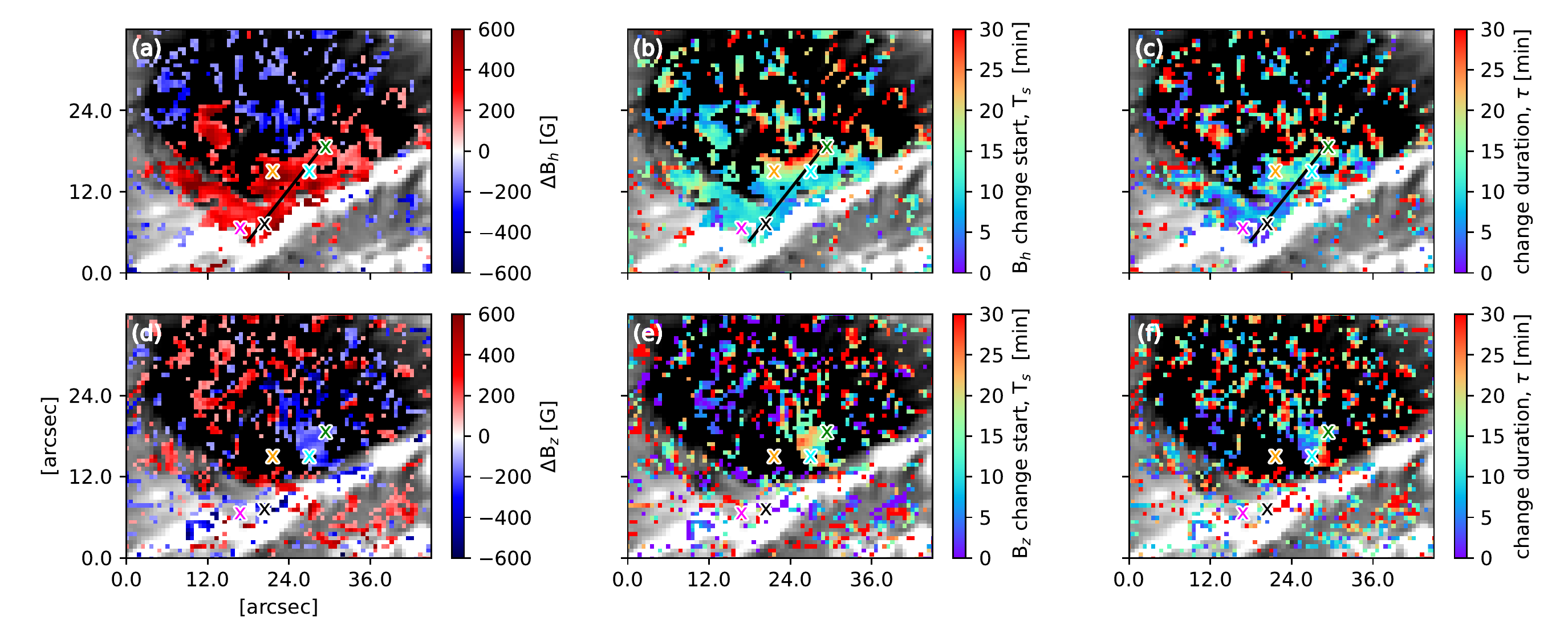}
    \includegraphics[clip,trim=0.0cm  -0.3cm 0cm 0cm, width=0.24\linewidth]{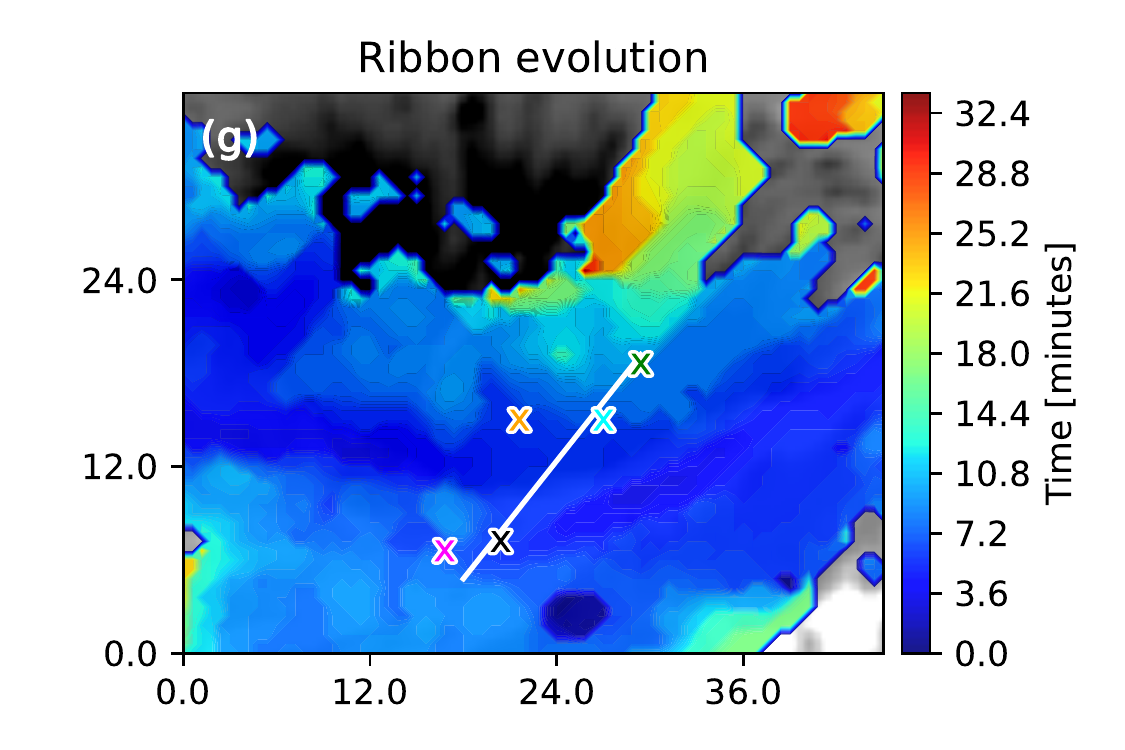}
    
    \caption{Characteristics of the changes in the field derived from the step-like function during the X2.2 flare on 2011 February 15 starting at 01:44 UT. \textit{Top panels}: the distribution of the changes in the magnetic field (a; $\Delta B_h$), the start time of the change in the field (b; T$_s$), and the duration of changes in the field (c; $\tau$) for the B$_h$ component. \textit{Bottom panels} (d--f): same as the top panels but for the B$_z$ component. The zero value in the T$_s$ is the flare start time. The right panel (g) shows the temporal evolution of ribbons, where color shows the initial time of ribbon brightening in each pixel. The spatial distribution of parameters along a black (white) line is shown in Figure \ref{fig:line_parameters}. The temporal evolution of pixels marked by cross symbols is shown in Figure~\ref{fig:step_fit_cases}. The background image in all panels shows the $B_z$ component of the magnetic field (saturated at $\pm800$~G), where white and black refer to the positive and negative polarities, respectively.}
    \label{fig:fittep_maps}
\end{figure*}

\begin{figure*}
    \centering
    \vspace{5mm}
    \includegraphics[width=1.\linewidth]{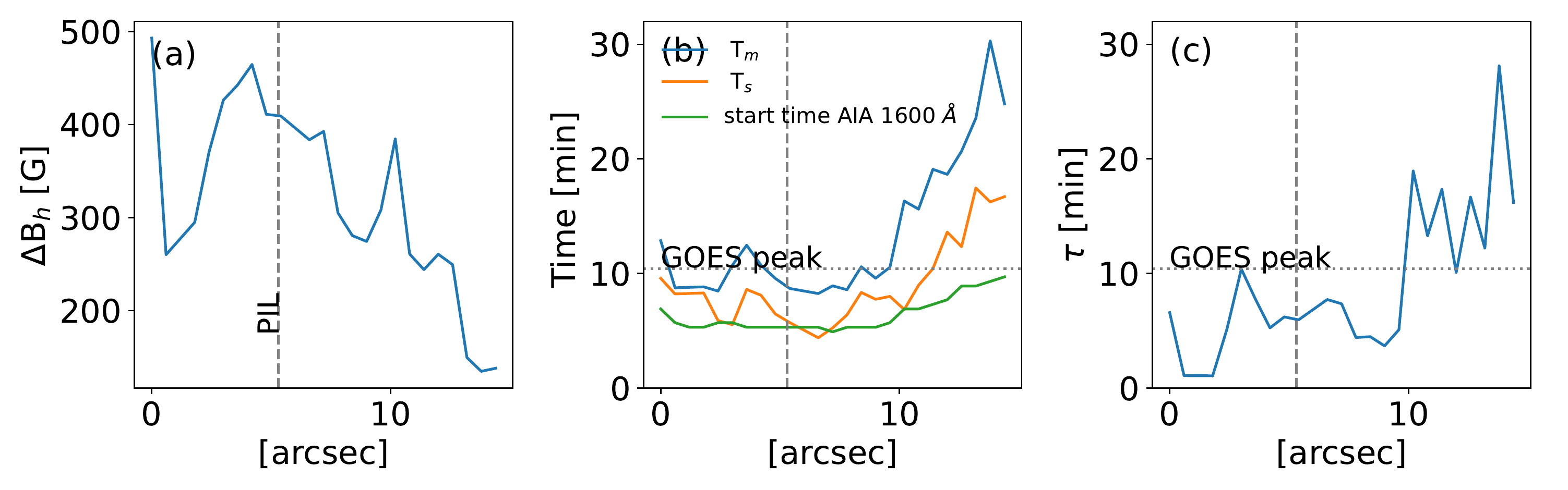}
    \caption{Properties of the changes in the field along a solid line highlighted in Figure \ref{fig:fittep_maps}. (a) Change in $\Delta B_h$; (b) mid-time of the change in the field (T$_m$; blue line), start time of the change in the field (T$_s$; orange line), and the start time of the brightening of ribbons (AIA 1600~\AA; green line); and (c) duration of the field change $\tau$. The zero value in time (panel b) is the start time of the flare. Dotted horizontal and dashed vertical lines denote the GOES flare peak time and the PIL location, respectively.}
    \label{fig:line_parameters}
\end{figure*}

\begin{figure}
    \centering
    \includegraphics[width=1\linewidth]{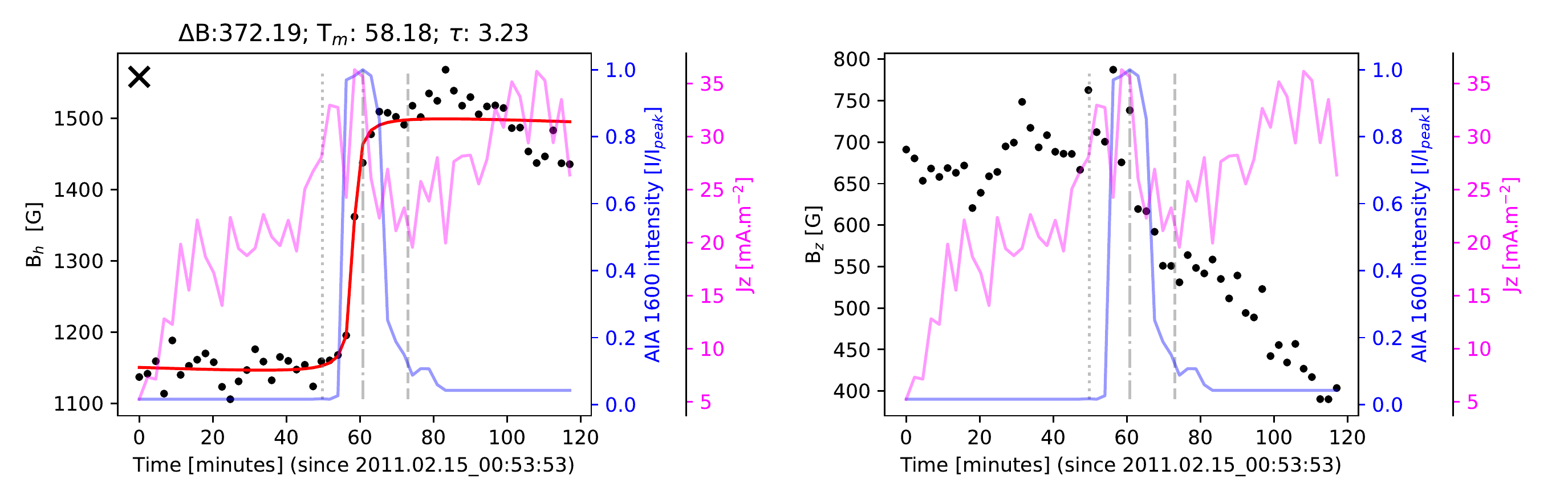}
    \includegraphics[width=1\linewidth]{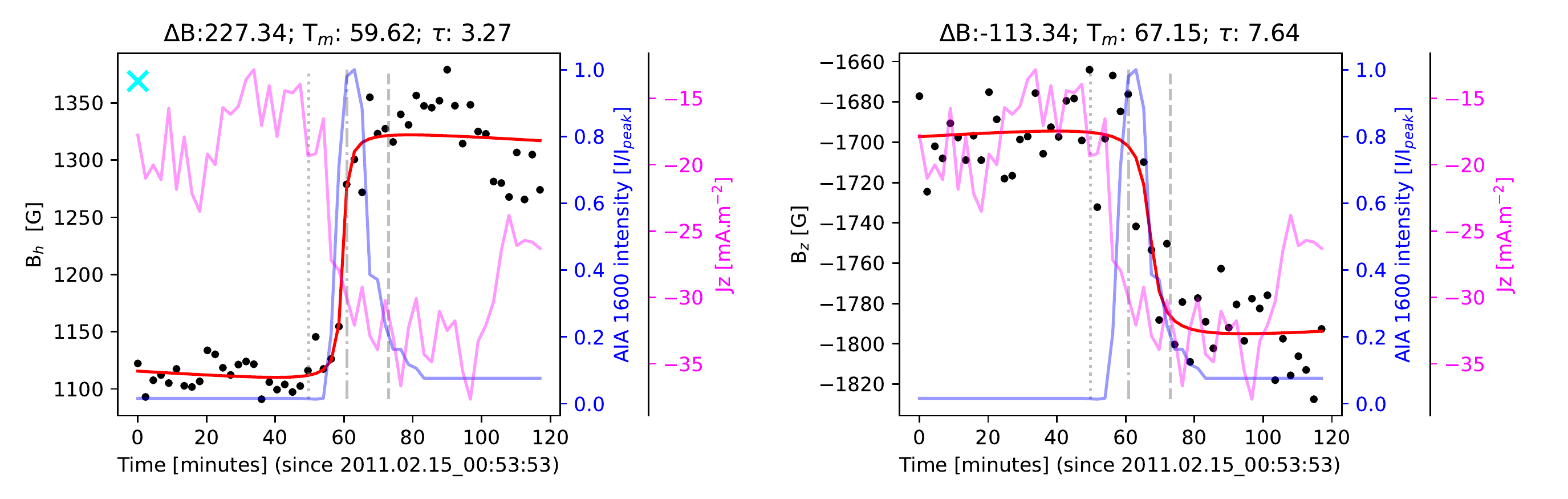}
    \includegraphics[width=1\linewidth]{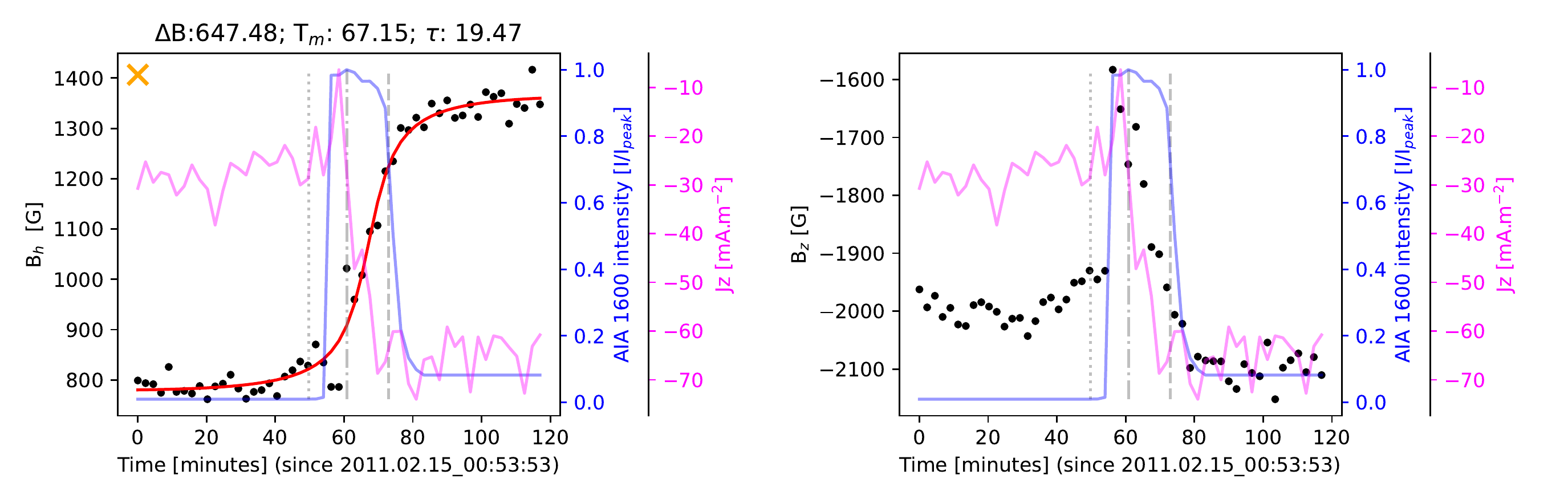}
    \includegraphics[width=1\linewidth]{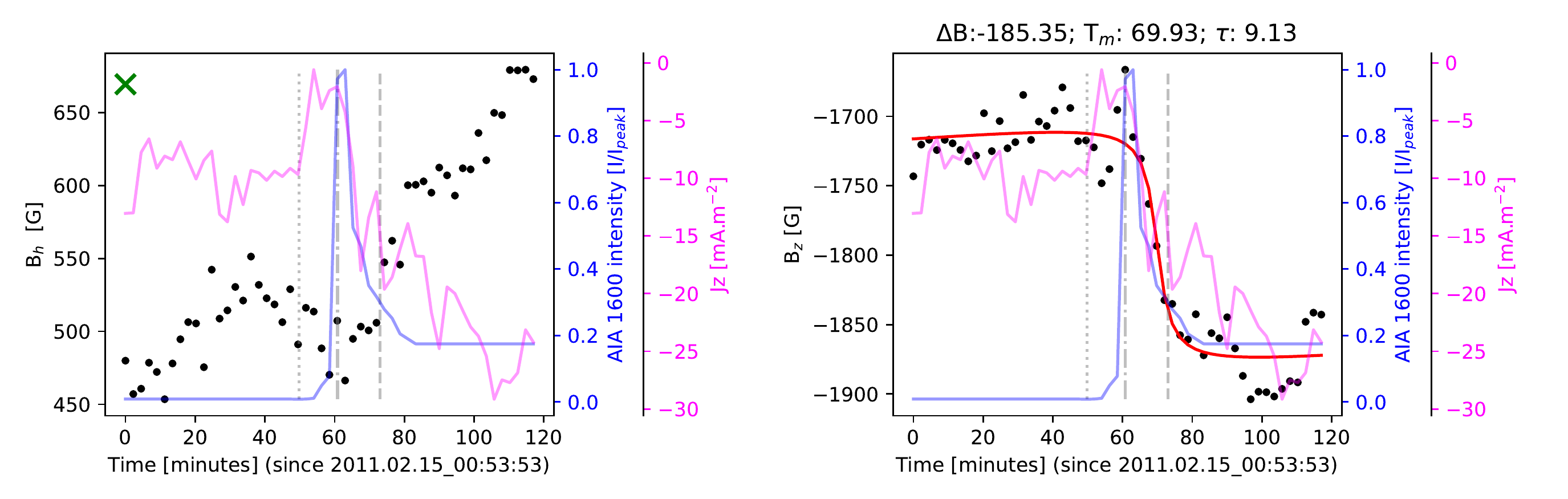}
    \caption{Temporal evolution of $B_z$, $B_h$, $J_z$, and the intensity of AIA 1600~\AA\ in four pixels highlighted by colored cross symbols in Figure \ref{fig:fittep_maps}. The black dot refers to the observed $B_z$, $B_h$ values, whereas the best-fitted profiles obtained from the step-like function are represented by the red solid line. The blue and pink lines refer to the intensity of AIA~1600~\AA\ and the vertical current density (J$_z$), respectively. The vertical dotted, dashed-dotted, and dashed line indicates the GOES flare start, peak, and end times, respectively.}
    \label{fig:step_fit_cases}
\end{figure}

\begin{figure}
    \centering
    \includegraphics[width=1\linewidth]{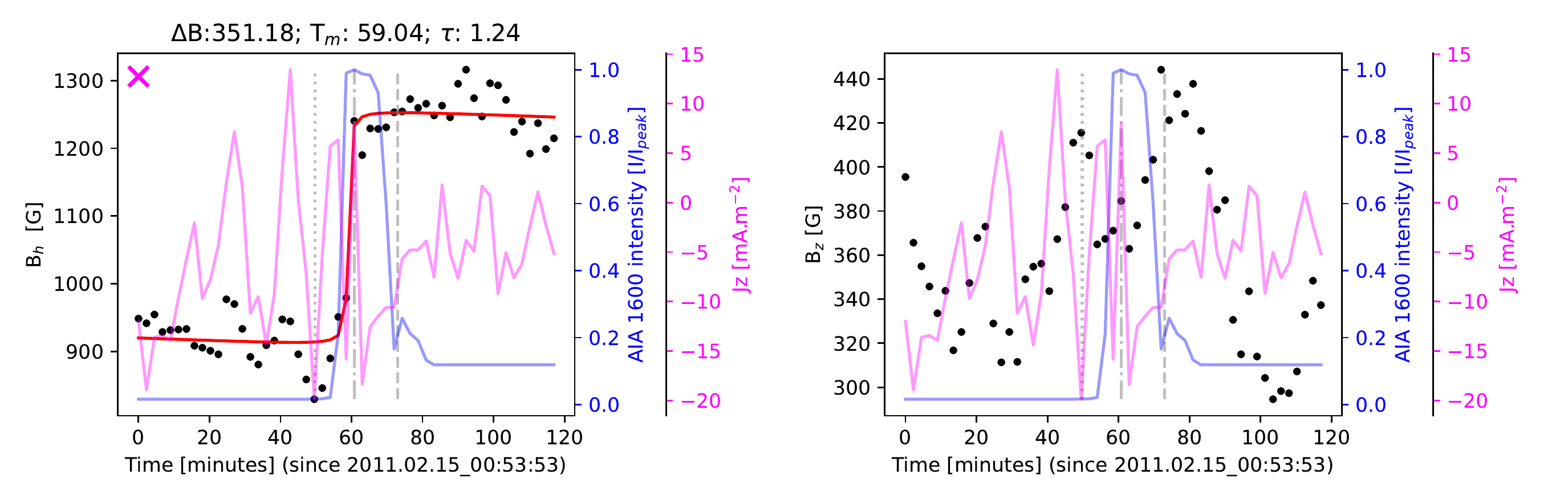}
    \caption{Same as Figure~\ref{fig:step_fit_cases}, but for the pixel highlighted by the magenta cross symbol in Figure \ref{fig:fittep_maps}.}
    \label{fig:step_fit_cases_extra}
\end{figure}

In the following sections, we first describe the characteristics of the changes in the field and their association with UV emission (e.g., an X2.2 flare observed on 2011 February 15). We then summarize the statistical results and the correlations between different variables derived from the step-like function for $37$ events. 

\subsection{Case Study: Analysis of Field changes in the X2.2 flare on 2011 February 15}
As an example, we first analyze and derive the characteristics of field changes in an X2.2-class flare (SOL2011-02-15T01:56). The temporal evolution of flare ribbons associated with this flare is shown in Figure \ref{fig:ribbon_evolution}, where the violet and red colors correspond to the early and late stages of the flare, respectively. Similar to the method used in \cite{2017ApJ...845...49K}, the total unsigned ribbon area is estimated using the cumulative ribbon pixels. The evolution of the
magnetic fluxes swept up by ribbons in positive ($\Phi^+_\mathrm{ribbons}$) and negative ($\Phi^-_\mathrm{ribbons}$) polarities, shown in the bottom panel of Figure \ref{fig:ribbon_evolution}, are determined using the Equation~3 given in \cite{2017ApJ...845...49K}.

\subsubsection{Magnetic Field Changes During an X2.2 Flare}
Figure \ref{fig:fittep_maps} shows the characteristics of the field change during an X2.2-class flare (SOL2011-02-15T01:56): field change magnitude $\Delta B_h$, field change start time $T_s$, and duration $\tau$ obtained from the step-like function (see Equation \ref{eq_stepfit}) for the $B_h$ (Figure \ref{fig:fittep_maps}; a--c ) and $B_z$ (Figure \ref{fig:fittep_maps}; d--f ) components of the magnetic field. Additionally, panel~g shows the temporal evolution of chromospheric ribbons, where color corresponds to the initial ribbon brightening in each pixel. As described in Section~\ref{sec_stepfit}, we fitted the temporal evolution of $B_h$ and $B_z$, separately. In total, $2676$ out of 6195 pixels satisfied our criteria (see Sec.~\ref{sec:criteria field change}) for $B_h$.  As demonstrated in the maps, pixels showing $\Delta B_{h}$ and $\Delta B_{z}$ are located in the umbra, penumbra, and near the PIL of the AR. However, the majority of pixels showing $\Delta B_h$ are located near the PIL, which is in agreement with previous studies (e.g., \citealt{2017ApJ...839...67S, 2022ApJ...934L..33L}). Figure \ref{fig:fittep_maps} also demonstrates that, in comparison to $\Delta B_h$, $\Delta B_z$ is less pronounced and is distributed in small patches of pixels over the whole AR. Panels b and c illustrate how the start time of the change in the field and the duration of the change in the field are distributed over the AR. We find that the pixels located close to the PIL exhibit a fast and early permanent change in $B_h$, whereas the pixels located away from the PIL have slower- and later-occuring permanent changes in the field (i.e. larger $T_s$ and $\tau$). For the $B_z$ component, $\Delta B_z$, $T_s$, and $\tau$ are sparsely distributed over the AR without a clear pattern. 

 In Figure \ref{fig:line_parameters} we demonstrate how the B$_h$, $\tau$, T$_m$, T$_s$, and temporal/spatial evolution of ribbons change along a solid line shown in Figure \ref{fig:fittep_maps}.
 The pixels located near the PIL show a strong $\Delta$B$_h$ value, but the magnitude decreases for the pixels located away from the PIL. Moreover, the T$_m$ and T$_s$ parameters increase gradually as we go further away from the PIL. Pixels located near the PIL exhibit a fast and early change relative to the GOES peak time, whereas the pixels located away from the PIL have longer $\tau$, $T_s$, and $T_m$ values. Panel b also shows that ribbons appear earlier than T$_m$ and T$_s$ before the GOES peak time. In summary, Figures \ref{fig:fittep_maps} and \ref{fig:line_parameters} suggest that pixels located near the PIL exhibit early changes in the field and shorter duration of the  changes with larger magnitudes of $\Delta$B$_h$ compared to pixels located $\sim$10\arcsec~from the PIL, which is in line with previous studies \citep{2017ApJ...839...67S,2018ApJ...852...25C, 2022ApJ...934L..33L}.

In Figure \ref{fig:step_fit_cases}, we highlight the temporal evolution of four pixels, marked with colored cross symbols in  Figure \ref{fig:fittep_maps}. 
For these four pixels, we show the temporal evolution of the horizontal and vertical components of the magnetic field, the intensity of AIA~1600~\AA\ and the vertical current density ($J_z = \mu_0^{-1} (\nabla \times B)$).
The temporal evolution of $B_{h,z}$ shows significant evidence of a change in the field. 
Some of the pixels show a clear and fast ($\tau = 3.23$~min) permanent step-like change in $B_{h}$, whereas some pixels exhibit a slower ($\tau = 19.47$ minutes) and larger change in $\Delta B_h$ ($\sim$ $647$~G). In some pixels, we noticed a permanent step-like change in both the $B_{h}$ and $B_{z}$ components of the magnetic field, although generally, $\Delta B_h$ is stronger than $\Delta B_z$. Moreover, we also found pixels exhibiting no clear step-like change in the $B_h$, but a clear step-like change pattern in $B_z$. 

There are some cases where the magnetic field vector retrieved from the inversions is not reliable as the Stokes observations are impacted by flare. As an example, the pixel marked by the orange-colored cross symbol (see Figure \ref{fig:step_fit_cases}) shows a sudden abrupt/transient change in $B_z$. For this pixel, the temporal evolution of $B_z$ shows a transient change from $-$2000~G (pre-flare) to $-$1600~G (around the flare peak time). This transient change of 400~G is likely produced by the flare emission. In the pixels, located at flare emission sites, the inferred magnetic field has more uncertainties due to poor fitting of flare-distorted Stokes profiles under Milne-Eddington approximation (VFISV; \citealt{2011SoPh..273..267B, 2017ApJ...839...67S}). The distortion in the Stokes profiles during this X2.2-class flare (SOL2011-02-15T01:56) has been reported in previous studies \citep{2012ApJ...747..134M, 2014RAA....14..207R, 2017ApJ...839...67S}.
Given the nature of the step-like function, the Equation~\ref{eq_stepfit} cannot fit pixels exhibiting a transient change in $B_z$ or $B_h$. Consequently, we get a chi-square value above the threshold limit (see Section~\ref{sec:criteria field change}). Such pixels are excluded from our analysis based on the chi-square value obtained after fitting the time series with Equation~\ref{eq_stepfit}, but need more attention to understand the flare-related artifacts in the Stokes profile, which is beyond the scope of this study.

In addition to the temporal evolution of $B_h$ and $B_z$, Figure~\ref{fig:step_fit_cases} also shows co-temporal and co-spatial $J_z$ and the intensity of AIA~1600~\AA{} for the selected pixels. To investigate the relation between $J_z$ and $B_{h}$, we measure the change in $J_z$ at $T_s$ and $T_e$, $\Delta J_z = J_z(T_e) - J_z(T_s)$, for all pixels showing a step-like change. From the comparison of $\Delta J_z$ and $\Delta B_{h}$ we find that they are not related to each other. As illustrated in Figure~\ref{fig:step_fit_cases}, $J_z$ shows some remarkable step-like behavior near the PIL, but these patterns are not consistent with the permanent change in the field (e.g., top panels of Figure~\ref{fig:step_fit_cases}).
From the analysis of six major flares, \citet{2012ApJ...759...50P} also reported that changes in $J_z$ show no consistent patterns. One of the possibilities for this behavior could be the nature of $J_z$, which is derived from derivatives of the horizontal components ($B_x$ and $B_y)$ of the magnetic field.

During our analysis, we also found that there are some pixels ($\sim$1\% of total pixels) exhibiting fast clear step-like changes ($>$200~G) in the $B_h$ component of the magnetic field. As an example, Figure~\ref{fig:step_fit_cases_extra} shows that the duration of the change in the  magnetic field (1.24~minutes) in the $B_h$ component obtained after fitting with Eq~\ref{eq_stepfit} is less than the HMI cadence (2.25~min). It also exhibits a clear step-like change in the $B_h$ (351~G).
As any change below the cadence of HMI magnetograms would be less reliable, we have neglected pixels exhibiting a duration of a change in the field below the HMI cadence.

\subsubsection{Relationship between the AIA 1600~\AA{} Emission and the Magnetic Field Changes During an X2.2 Flare}
For the analyzed X2.2 flare occuring on 2011 February 15, the change in the intensity of AIA~1600~\AA{} generally starts after the GOES start time of the flare, peaks at the GOES flare peak time, and then decreases gradually (see Figure~\ref{fig:step_fit_cases}, blue lines). We also analyzed how the selected pixels exhibiting permanent changes in the field are related to UV emission (AIA 1600~\AA). We find that some pixels exhibiting UV brightness show a permanent change in B$_h$, but not all brightening pixels are accompanied by change in the field, which is in line with \cite{2012ApJ...760...29J}. For the X2.2 flare out of 2676 pixels showing a permanent field change, only 41$\%$ of pixels accompanied the UV brightening. 
 
\begin{figure}
    \centering
    \includegraphics[width=1\linewidth]{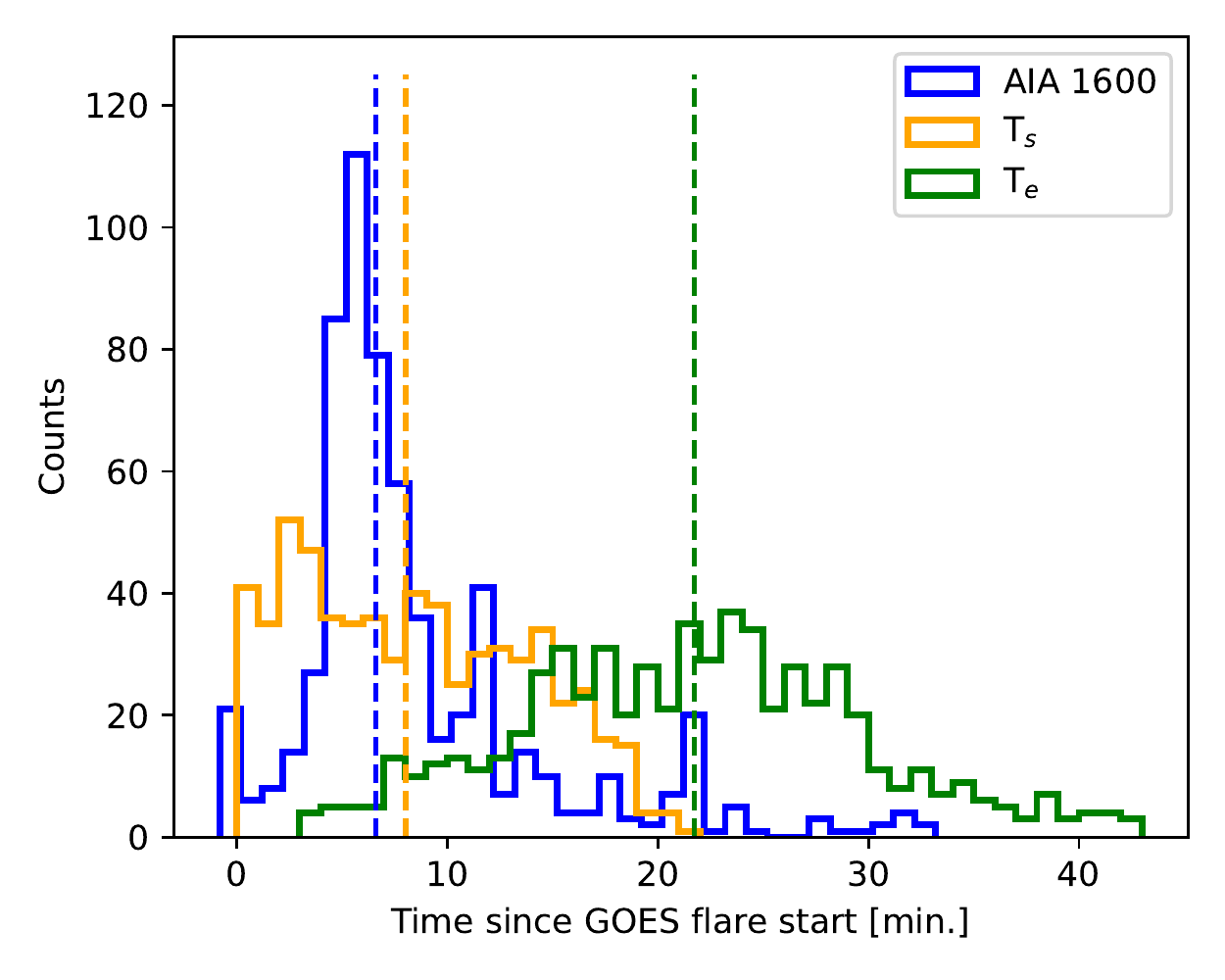}
    \caption{Histograms of the start time of the change in the intensity of AIA 1600~\AA{}, the start time of the change in the $B_h$ (T$_s$), and the end time (T$_e$) for the SOL2011-02-15T01:56 X2.2-class flare. The vertical dashed lines denote the median values of the start time of the intensity of AIA 1600~\AA~(blue), the start (orange) and the end (green) times of the change in the B$_h$.}
    \label{fig:aia_b_change}
\end{figure}

In Figure~\ref{fig:aia_b_change} we show the relation between the start time of the UV brightening and permanent field change in $B_h$. To determine the start time of the UV brightening we employed an intensity threshold that is 3 times larger than the median value of the quiet-Sun intensity of AIA~1600~\AA{}. We find that the majority of pixels start exhibiting a change in the intensity of AIA~1600~\AA{} after the GOES flare start time. However, $2.6\%$ pixels out of 1099 show a rise in  intensity before the GOES flare start time. These early UV brightenings can be caused by sequential chromospheric brightenings that could be caused by enhanced particle beams from the corona \citep{2005ApJ...630.1160B}. In most of the pixels, the UV emission starts early compared to the field change start time. The median values of the start time of the change in the intensity of AIA, $T_s$, and $T_e$ are 6.6, 8.1, and 21.7 minutes, respectively. The median time delay between the start of the UV emission and the start of the change in the magnetic field is 1.5 minutes.

We applied the same procedure to the remaining events to determine the delay in the changes in the magnetic field associated with flare ribbons. In $\sim$85\% of events, we find that the UV emission starts early compared to the start time, $T_s$, of the change in the field. For these events, the median value of the delay of the change in the field is 4.4 minutes and the maximum delay is around 19 minutes. On the other hand, $\sim$15\% of events showing early $T_s$ relative to start time of the change in the intensity of AIA could be due to irregular small-scale brightening, which is not detected by our intensity threshold.

\subsection{Statistical Properties of Field Changes in 37 Flares}
\begin{figure}
    \centering
    \includegraphics[width=1\linewidth]{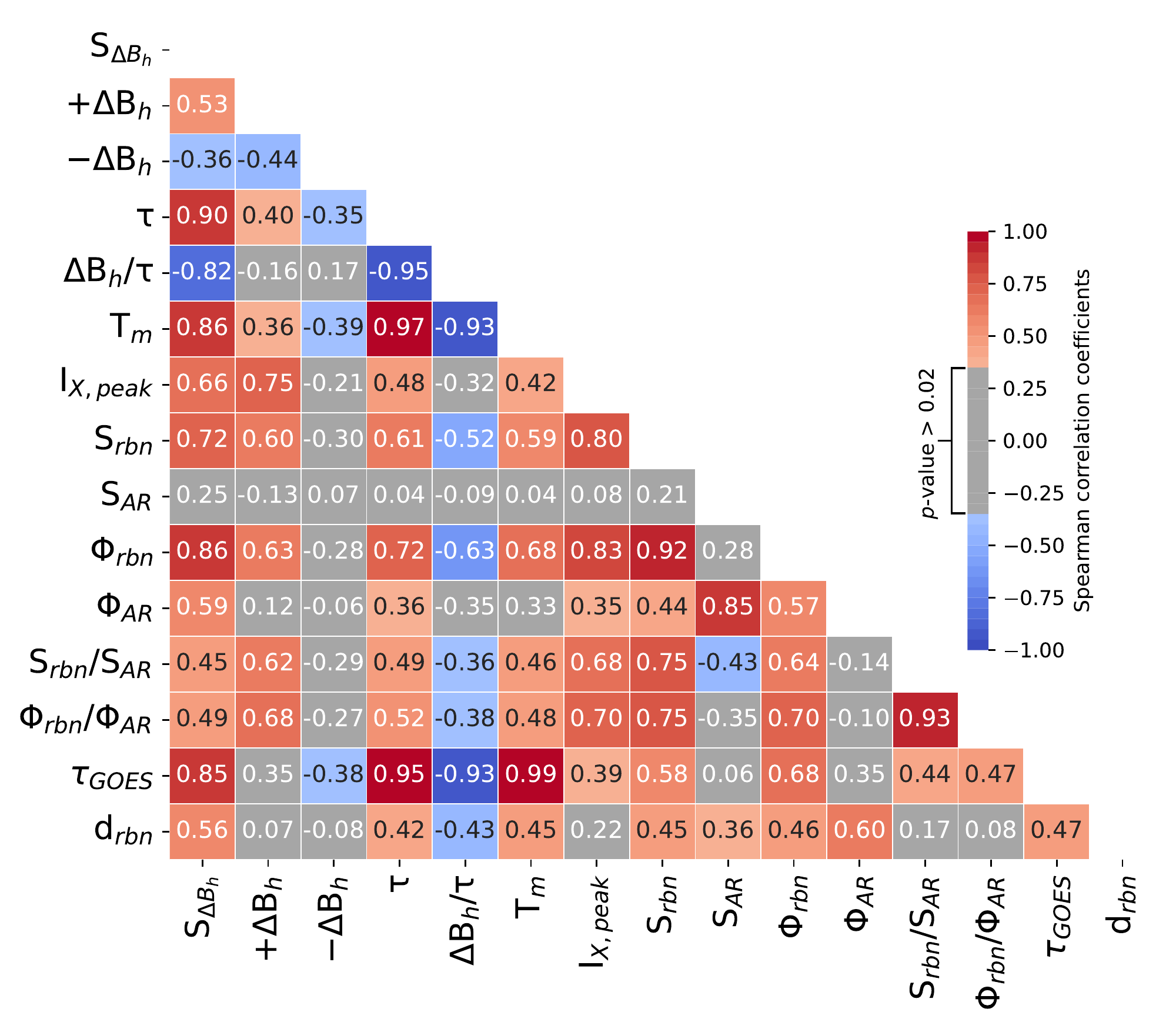}
    \caption{Correlation matrix showing Spearman correlation coefficients (cc) between different properties of B$_h$ for 37 flares. The correlations are performed among the total area showing step-like change (S$_\mathrm{\Delta B_h}$), positive and negative change in B$_h$, duration of the field change ($\tau$), change rate of the field ($\Delta$B$_h$/$\tau$), mid-time of the change in the field (T$_m$), GOES X-ray peak flux (I$_{X, peak}$), total area of ribbons (S$_\mathrm{rbn}$), active region area (S$_{AR}$), total magnetic flux in ribbons ($\Phi_\mathrm{rbn}$), total magnetic flux in AR ($\Phi_{AR}$), ratio of ribbon to active region area (S$_\mathrm{rbn}$/S$_{AR}$), ratio of ribbon flux to active region flux ($\Phi_\mathrm{rbn}$/$\Phi_{AR}$) , duration of the GOES flare ($\tau_{GOES}$), and ribbon distance (d$_\mathrm{rbn}$). All Spearman correlation coefficient values above $|0.35|$ have $p$-values $<0.02$ and are statistically significant. The statistically insignificant ($p$-values $>0.02$) correlation coefficients are highlighted in gray color, whereas the remaining blue/red colors indicate statistically
significant values.}
    \label{fig:correlation_matrix}
\end{figure}

\begin{figure*}
    \centering
    \includegraphics[width=1\linewidth]{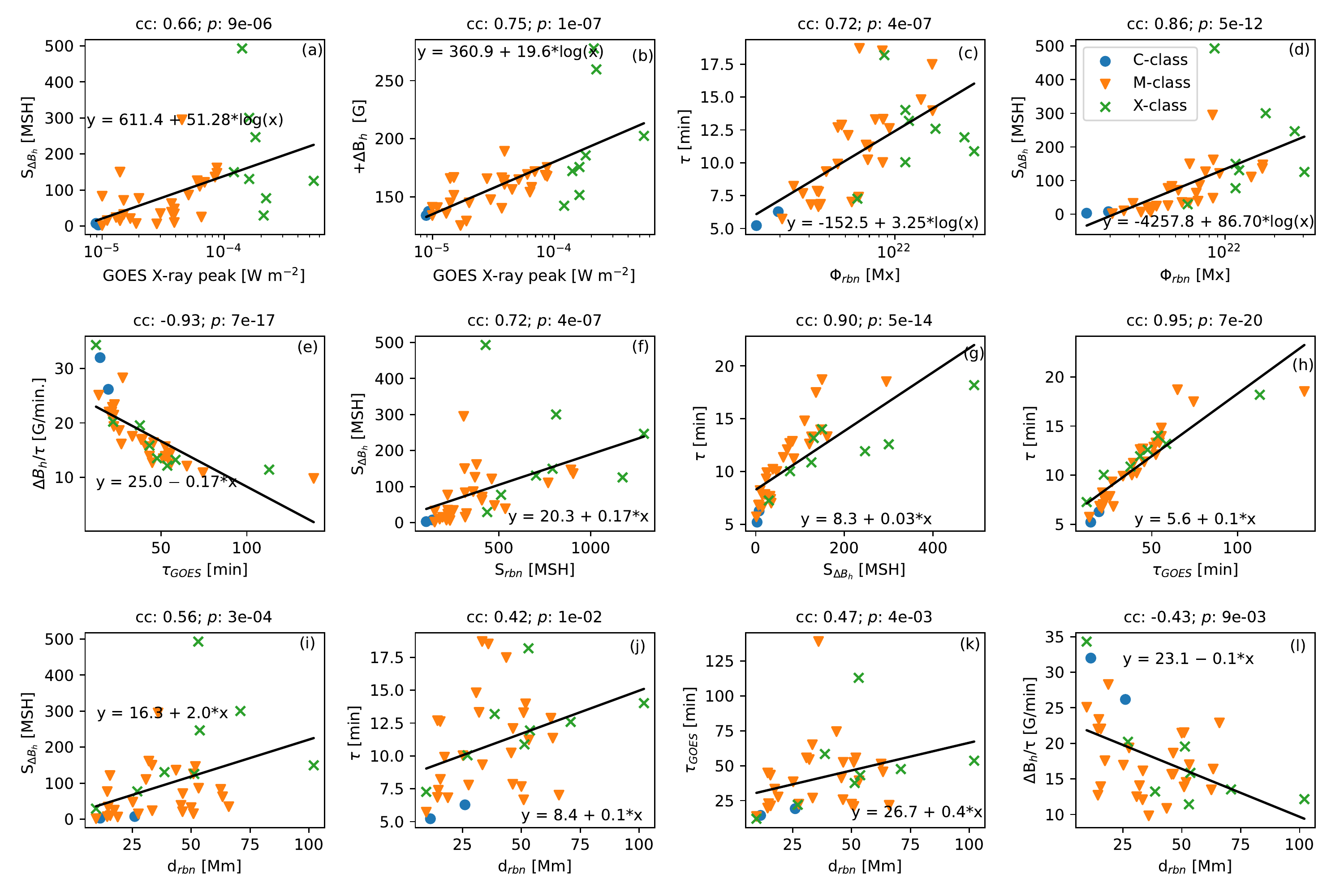}
    \caption{Statistical properties of the changes in the horizontal field during 37 flares. Scatter plots showing the relation between (a) GOES X-ray flux vs. total area showing permanent changes (S$_\mathrm{\Delta B_h}$), (b) GOES X-ray flux vs. positive permanent change in B$_h$ ($+\Delta B_h$), (c) total ribbon flux ($\Phi_\mathrm{rbn}$) vs. duration of the change in the field ($\tau$), (d) $\Phi_\mathrm{rbn}$ vs. area $\Delta B_h$, (e) duration of the GOES flare ($\tau_{GOES}$) vs. the rate of the change in the field ($\Delta B_h/\tau$), (f) total area of ribbons (A$_\mathrm{rbn}$) vs. area $\Delta B_h$, (g) area $\Delta B_h$ vs. $\tau$, (f) T$_{GOES}$ vs. $\tau$, (i) ribbon distance (d$_\mathrm{rbn}$) vs. area $\Delta B_h$, (j) d$_\mathrm{rbn}$ vs. $\tau$, (k) d$_\mathrm{rbn}$ vs. $\tau_{GOES}$, and (l) d$_\mathrm{rbn}$ vs. $\Delta B_h/\tau$. In the top panel, the abscissa is in log scale. The solid line refers to the linear fit between parameters. In each plot, `cc' and `$p$' correspond to the Spearman correlation coefficient value and Spearman coefficient $p$-value, respectively. In all cases, the correlation is statistically significant ($p$-value $\le$0.01). The filled circle, triangle, and cross symbols correspond to C-, M-, and X-class flares, respectively. The area is expressed in millionths of a solar hemisphere, which is equivalent to 3$\times$10$^6$ km$^2$.}
    \label{fig:scatter_plots}
\end{figure*}

In this section, we present the statistical analysis of all events shown in Table~\ref{tab_flareinfo}. For consistency, we employed the same procedure, as described above for all events. (see Section~ \ref{sec_stepfit}). Here we discuss the statistical properties of the physical parameters derived after fitting the time series of B$_h$ for each pixel with the step-like function (Equation~\ref{eq_stepfit}; see Table~\ref{tab:bh_change} and Figure~\ref{fig:appendix_overview_bh_change} in the Appendix). Additionally, we also analyzed the relationship between the properties of the change in the field and other AR and flare parameters including the intensity of the GOES X-ray peak flux (I$_{x,peak}$), S$_{AR}$, S$_\mathrm{rbn}$, $\Phi_{AR}$, $\Phi_\mathrm{rbn}$, $\tau_{GOES}$, and d$_\mathrm{rbn}$. 

Figure \ref{fig:correlation_matrix} shows the Spearman correlation coefficient (cc) between different parameters derived from all flaring events. The strength of the correlation is color coded. To describe the qualitative strength of the correlation we adopted the following guideline given by \cite{2017ApJ...845...49K}: cc $\in$ [0.2, 0.39]---weak, cc $\in$ [0.4, 0.59]---moderate, cc $\in$ [0.6, 0.79]---strong, and cc $\in$ [0.8, 1.0]---very strong. 

Figure \ref{fig:scatter_plots} shows examples of scatter plots between the derived parameters shown in Figure \ref{fig:correlation_matrix}. We find that the median $\Delta$B$_h$ value for all events ranges from 100--300~G (Figure~\ref{fig:scatter_plots}b). We also find that the total area showing a permanent change in B$_h$ and the magnitude of $\Delta$B$_h$ are strongly related with the GOES X-ray flux (Figure~\ref{fig:scatter_plots}a and b). This suggests that stronger flares affect larger areas in the photosphere, which is in agreement with previous studies \citep{2010ApJ...724.1218P, 2018ApJ...852...25C}. However, the duration of the change is only moderately related to the GOES X-ray flux. 

Although the characteristics of a change in a field, such as $\Delta$B$_h$, $\tau$, the rate of the change in the field ($\Delta$B$_h$/$\tau$) shows a weak or no relation with the AR area, they are very strongly related to the flare parameters (ribbon magnetic flux and flare-ribbon area). The duration of the flare, $\tau_{GOES}$, is positively correlated with the $S_\mathrm{rbn}$, $\Phi_\mathrm{rbn}$, and the class of flares (GOES X-ray peak flux). From the analysis of $2956$ flares, \citet{Reep2019} reported that in smaller flares the duration of the flare, defined as the FWHM of the GOES X-ray light curve ($\tau_\mathrm{FWHM}$), is not correlated with the ribbon area, $S_\mathrm{rbn}$  (cc=0.2, C-class). On the other hand, they found that the correlation increases for larger M- and X-class flares: cc=0.6 (M-class) and cc=0.9 (X-class). In a different study, \citet{2017ApJ...834...56T} analyzed $51$ large flares ($\ge$M5.0-class), finding that the $\tau_\mathrm{FWHM}$ is linearly correlated with $S_\mathrm{rbn}$, $\Phi_\mathrm{rbn}$, and ribbon separation, in agreement with our study. 


According to the standard flare model the ribbon separation, d$_\mathrm{rbn}$, generally refers to the footpoint separation of flare loops. If we assume that the flare loops are semicircular in shape, then the d$_\mathrm{rbn}$ is proportional to the height of the reconnecting loops and loop length. We find that $\tau_{GOES}$ is moderately correlated with d$_\mathrm{rbn}$ (cc=0.47). From the analysis of stronger flares (class M5.0 and above), \cite{2017ApJ...834...56T} found that the reconnection timescale is proportional to the loop length with a slightly higher correlation coefficient ($\propto d_\mathrm{rbn}$, cc=0.8). Consequently, longer d$_\mathrm{rbn}$ value would give rise to a longer duration of the flare, which is similar to our result (see Figure \ref{fig:scatter_plots}k). Using hydrodynamic modeling, \cite{2017ApJ...851....4R} also found a clear linear correlation between the ribbon separations and the FWHM of GOES light curves, indicating that the primary factors that control a large-flare timescale are the duration of the reconnection and the loop length. 

The rate of the change in the field, $\Delta B_h/\tau$, is inversely proportional to the ribbon separation d$_\mathrm{rbn}$ (Figure~\ref{fig:scatter_plots}) and the duration of the GOES flare $\tau_{GOES}$ (Figure~\ref{fig:scatter_plots}). Events having shorter $\tau$ values and smaller loop sizes exhibit faster changes in the field. This relation suggests that a low-lying loop or smaller d$_\mathrm{rbn}$, that is moderately correlated with $\tau$, would result in a fast $\Delta B_h/\tau$.
In our data set the median value of the duration of the change in the field, $\tau$, ranges from 5--18.7 minutes, where stronger flares exhibit longer $\tau$. 

Finally, the scatter plot between $\tau$ and $\tau_{GOES}$, which is defined as the time difference between the GOES flare start and end times, shows a remarkably strong linear relation with a Spearman correlation coefficient value of 0.95. We find that the duration of permanent changes in the field ranges from 23\%--42\% of the total duration of flaring, with an average value of 29\%.

\section{Discussion}
\label{sec: Discussion}
We present a statistical analysis of changes in the magnetic field associated with 37 flares. We investigate how the photospheric magnetic field vector changes using high-cadence vector magnetograms obtained from the HMI/SDO. We also examine how the characteristics of the change in the field are associated with the ribbon morphology and the UV emission. Although there are different types of changes in the field in the B$_h$ and B$_z$, we focus on the step-like and permanent changes in $B_h$. The characteristics of the change in field are obtained by fitting a time series of each pixel by a step-like function. 

The increase in B$_h$ or increase in field inclination in all events, mainly near the PIL, is in agreement with previous studies \citep{2017ApJ...839...67S,2019ApJS..240...11P, 2022ApJ...934L..33L}. The high-resolution vector magnetograms from HMI/SDO allowed us to investigate the fast permanent change in the field ($>$135~s). Nevertheless, we find pixels showing permanent changes on faster time scales than the cadence of our data set. This suggests that further high-cadence observations are needed to explore the relation of fast photospheric changes with flares. We also noticed that a permanent change in the field or a step-like change is also evident in the temporal evolution of B$_z$, but these changes are scarcely distributed over the FOV compared to the B$_h$. Due to the lack of statistics, we did not analyze the change in B$_z$ in detail. 

It is well known that a flare can occur anywhere in the upper atmosphere. If a loop is considered to be a semicircular shape, then the ribbon distance, d$_\mathrm{rbn}$, would be proportional to loop length and the reconnection height \citep{2017ApJ...834...56T}. Consequently, a smaller d$_\mathrm{rbn}$ would correspond to larger energy release in the lower atmosphere where fields are stronger, whereas a larger d$_\mathrm{rbn}$ (longer loops) would correspond to smaller energy release in the higher layers of the solar atmosphere. Based on the above assumptions, we can speculate that a flare with smaller d$_\mathrm{rbn}$ would release energy in the deeper layers of the solar atmosphere and lead to larger and faster changes in the magnetic field. Recently, \cite{2022ApJ...934L..33L} reported that the initial ribbon separation is roughly inversely proportional to the mean value of the change in $B_h$ in $35$ solar flares (21 X- and 14 $>M6$-class flares), especially for smaller distances (cc = $-$0.4). In contrast, using a much weaker flare sample of 8 X-, 7 $>M6$ and 22 $<M6$ flares, here we find that magnitude of the change in positive $B_h$ shows no correlation with the ribbon distance (cc = 0.07). This difference could be due to, e.g. our different approach to estimating the ribbon separation or a different physical process at play for weaker flares that we analyze here.

\begin{figure*}
    \centering
    \includegraphics[width=0.9\linewidth]{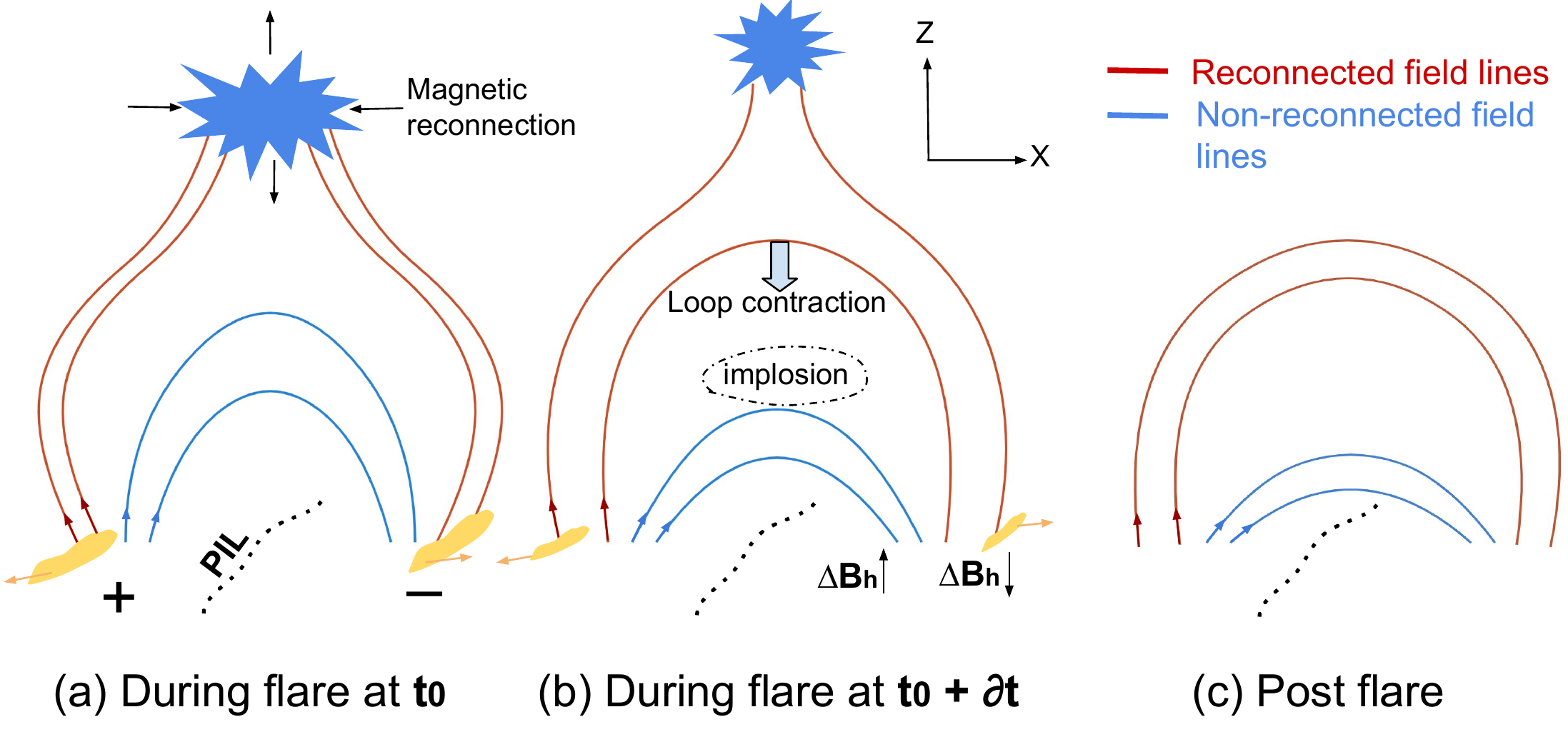}
    \caption{A sketch of the magnetic field configuration (a, b) during and (c) after the time of flaring. Solid lines refer to the field lines connecting opposite polarities, which are indicated by `$+$' and `$-$' signs. A dotted line refers to the PIL, whereas the ribbons are shown with a yellow-colored area near the footpoints. Blue field lines contract due to coronal implosion whereas the newly reconnected field lines shown in red shrink after reconnection. Here, the ``Z" axis is perpendicular to the solar surface, whereas ``X" is along the solar surface.}
    \label{fig:pre_post_cartoon}
\end{figure*}

Figures \ref{fig:correlation_matrix} \& \ref{fig:scatter_plots} illustrate that the area affected by a flare or pixels exhibiting a permanent change in the horizontal field, $S_\mathrm{\Delta B_h}$, is not only strongly correlated with d$_\mathrm{rbn}$ (cc = 0.56, Panel i) but also strongly correlated with the $\Phi_\mathrm{rbn}$ (cc = 0.86, Panel d), $S_\mathrm{rbn}$ (cc = 0.72, Panel f), $I_{X, peak}$ (cc = 0.66) and $\tau_{GOES}$ (cc = 0.85). We can speculate that a stronger flare, having larger $S_\mathrm{rbn}$, $\Phi_\mathrm{rbn}$, and $\tau_{GOES}$, would give rise to longer reconnection processes, and thus affect more pixels with a longer duration of a change in the field. Consequently, a stronger flare with a larger ribbon area and magnetic flux can penetrate and affect the deeper layers, resulting in an increase in $B_h$ mainly near the PIL. Furthermore, no clear relation between the d$_\mathrm{rbn}$ and the magnitude of the change in the field $\Delta B_h$ (cc = 0.07) suggests that the magnetic field changes in the photosphere are not related to the reconnection heights but depend strongly on the $S_\mathrm{rbn}$ (cc = 0.60),  $\Phi_\mathrm{rbn}$ (cc = 0.63), and $I_{X, peak}$ (cc = 0.75).

The observed increase in $B_h$ could related to the coronal implosion conjecture, where the coronal field lines contract after releasing stored magnetic field energy \citep{2000ApJ...531L..75H}. In this conjecture, the loop contraction arises due to a decrease in magnetic pressure and volume reduction at the reconnection sites. Moreover, the velocity disturbances generated at the flare site could also be responsible for loop contraction without being reconnected \citep{2017ApJ...837..115Z, 2017ApJ...851..120S}. Additionally, an increase in $B_h$ could be a result of the reconnection-driven contraction of sheared flare loops, as has been recently demonstrated by \cite{2019ApJ...877...67B} using a 3D magnetohydrodynamic simulation. During magnetic reconnection, magnetic field lines of opposite direction break and then reconnect, forming newly reconnected field lines that contract and accelerate plasma away from the reconnection site \citep{Longcope2009}. The newly formed field lines contract toward the deeper layers to attain a stable configuration or relax to a quasi-force-free state.

We suggest that the abovementioned mechanisms can all contribute to the observed change in $B_h$ during flares. A possible scenario demonstrating how the magnetic field structure changes during a flare is shown in Figure \ref{fig:pre_post_cartoon}. The sketch depicts that the inner loops lying between yellow flare ribbons might become more horizontal due to loop contraction caused by the magnetic implosion or velocity disturbances \citep{2017ApJ...851..120S}, whereas the field lines closer to outer loops and within the flare ribbons  might become more vertical from contracting reconnected loops  due to rearrangement of post-reconnection field lines following magnetic reconnection. We would like to note though that there is a large variation in the relationship between ribbons' morphology and locations of horizontal field increases, indicating that the change in the field is a result of both processes.

We also investigated how a permanent field change in the photosphere is associated with the UV emission. We find that not all pixels showing a permanent change in the field in $B_h$ are associated with the UV emission (enhancement of the intensity of AIA 1600~\AA). 
For all events, the percentage of pixels showing both an enhancement of the intensity of AIA 1600~\AA~and permanent changes in the field range from 4--50\%. 
As the magnetic field measurements in our data have higher noise, small changes ($<100~G$) in $B_{h,z}$ associated with ribbons are not analyzed in this study.

We also investigated how the start time of the change in the field is related to the start time of the UV emission for the pixels showing both UV and a permanent change in B$_h$. In 31 events, out of a total of 37 events, we find that the UV emission starts early compared to the start time of the change in the field, where the median and maximum delay are 4.4 and 19 minutes, respectively. This relation suggests that the majority of magnetic field changes in the photosphere are consequences of flares. From the analysis of four X-class solar flares, \citet{2012ApJ...760...29J} reported that the UV emissions preceded the photospheric changes in the field by 4 minutes on average with the longest lead being 9 minutes, which is in agreement with our findings. 

On the other hand, six events show early (a few minutes) changes in the field compared to UV emission. The reason for this early change is not clear. One of the possibilities to explain this could be the magnetic reconnection at deeper layers, whereas the UV emission would be a result of reconnection at higher layers or delayed particle acceleration. Recently, \cite{2015ApJ...806..173B} studied the correlation between abrupt permanent changes in the magnetic field and hard X-ray emission observed by RHESSI during six X-class flares. They also reported that the amplitudes of the change in the field peak a few minutes earlier than the peak of the hard X-ray signal. 

Why do the chromosphere brightenings show an early rise compared to the field changes in the photosphere? This could be related to the energy deposition by MHD or Alfv\'en waves generated by the sudden change in the field lines in the corona, though the real reason is not yet clear. These waves travel in all directions, including the lower solar atmosphere, and can take a few minutes to reach the bottom of the corona \citep{2008ASPC..383..221H}. The release of energy through the interaction of waves with the dense chromospheric plasma gives rise to chromospheric brightenings \citep{1982SoPh...80...99E, 2001ApJ...558..859D, 2008ApJ...675.1645F}. Consequently, we see chromospheric brightening first compared to the field changes in the photosphere.

We also find a remarkable positive correlation (cc~=~{0.95) between $\tau_{GOES}$ and $\tau$. This implies that a flare having a longer duration will result in a longer duration of changes in the field, irrespective of flare intensity. Although there are different types of changes in the field during a flare, on average 29\% of the total flare duration time exhibits permanent changes in the field. 

In this study we focused on the permanent step-like change; however, there are different types of changes occurring in the photosphere at different locations. One of our future studies will include an investigation of all kinds of changes in the field and their preferred locations not only in the photosphere but also in the chromosphere. Machine-learning algorithms would be useful to classify and identify different complex types of changes in the field occurring during a flare and thus could improve our understanding of magnetic imprints of flares in the lower solar atmosphere.

\section{Conclusion}
\label{sec: conclusion}
In this paper, we have utilized high-cadence vector magnetograms observed by HMI/SDO to investigate magnetic imprints in the photosphere during 37 flares, mostly M- and X-class, and their association with the ribbon morphology.
Our main findings are shown in Figures \ref{fig:correlation_matrix} and \ref{fig:scatter_plots} and are highlighted below.
\begin{enumerate}

    \item In all events, the pixels showing a permanent and step-like change in the horizontal $B_h$ and vertical $B_z$ components of the magnetic field are distributed all over the AR, but the majority of them are located close to the PIL for B$_h$. Pixels showing changes in  B$_z$ are less pronounced and are distributed in small patches over the whole AR.
    In all cases, the magnitude of the change in the field in B$_h$ is stronger than in B$_z$.
    
    \item We find that pixels located near the PIL exhibit an early change in the field and a shorter duration of change with larger magnitudes of change in the field in $\Delta B_h$ compared to pixels located $\sim$10\arcsec\ away from the PIL.
    
    \item We find no clear relation between the temporal evolution of vertical current density $J_z$ and $B_{h}$ field components for the pixels exhibiting permanent and step-like changes. Some pixels near the PIL show step-like changes in $J_z$ but they are not consistent with the permanent changes in the magnetic field. 
    
    \item We find that not all pixels showing permanent changes in the field in $B_h$ are associated with the UV emission or vice versa. For all events, the percentage of pixels showing both an enhancement of the inetnsity of AIA 1600~\AA~and a permanent change in the field range from 4\%--50\% of all pixels in the selected regions. In 31 events out of a total 37, we find that the UV emission starts early compared to the start time of the change in the field, with the median and maximum delays of around 4.4 and 19 minutes, respectively. 
    
    \item The median changes in the magnitude of the magnetic field $\Delta$B$_h$ for all events ranges from 100--300~G. We find that the total area showing a permanent change in B$_h$ and the magnitude of $\Delta$B$_h$ are strongly correlated with the GOES peak X-ray flux. 
    
    \item The characteristics of a change in the field such as magnitude of a permanent change in a field ($\Delta$B$_h$), duration of the change in the field ($\tau$), the rate of the change in the field ($\Delta$B$_h$/$\tau$) show weak or no relation with the AR area, but are very strongly related to the flare parameters ($\Phi_\mathrm{rbn}$, $S_\mathrm{rbn}$, $\tau_{GOES}$).
    
    \item For the first time, we find that the duration of the permanent change in the field, $\tau$, is strongly correlated with the duration of the GOES flare ($\tau_{GOES}$; cc = 0.95). We find that this duration of the permanent change in the field ranges from 14--42\% of $\tau_{GOES}$, with an average value of 29\%. The median value of the changes in the field ranges from 5--18.7 minutes, with the changes in $\tau_{GOES}$ ranging from 12--138 minutes.
    
    \item Finally, we find that the total area showing the permanent change (S$_\mathrm{\Delta B_h}$), the duration of change in the field ($\tau$), and the GOES flare duration ($\tau_{GOES}$) are positively correlated with the ribbon distance ($d_\mathrm{rbn}$), whereas the magnitude of the change in the field is not correlated with $d_\mathrm{rbn}$.
\end{enumerate}

In Figure \ref{fig:pre_post_cartoon} we summarize the results of our analysis where magnetic field changes in the horizontal and vertical components in the photosphere are a consequence of magnetic reconnection and magnetic field implosion. As a result of this combination, we observe an increase in $B_h$ near the PIL and  decrease in $B_h$ away from the PIL. A real configuration of magnetic field lines during a flare would be more complex at different heights. Therefore, to present a clear picture we need multi-height spectropolarimetric observations, especially in the lower solar atmosphere (e.g. the Daniel K. Inoue Solar Telescope, DKIST, \citealt{Rast2020}).

\begin{acknowledgements}
We would like to thank the anonymous referee for the comments and suggestions. We thank Dr. Brian T. Welsch for reading the manuscript and pointing out a mistake in the flare end time.
We thank the HMI team for providing us with the vector magnetic field SDO/HMI data and the AIA team for providing us with the SDO/AIA data. We acknowledge support from NASA LWS NNH17ZDA001N, NASA LWS 80NSSC19K0070, 
NASA ECIP 80NSSC19K0910  and NSF CAREER award SPVKK1RC2MZ3 (R.Y. and M.D.K.) 
This research has made use of NASA’s Astrophysics Data System. We acknowledge the community effort devoted to the development of the following open-source packages that were used in this work: numpy (\url{numpy.org}), matplotlib (\url{matplotlib.org}) and SunPy (\url{sunpy.org}).
\end{acknowledgements}

\bibliographystyle{aa}
\bibliography{new-ref}  

\appendix

\begin{table}[!h]
    \centering
            \caption{List of the properties of the change in  B$_h$ for 37 flares (see Table~\ref{tab_flareinfo}). For each event the median value of the following parameters is listed: total area showing step-like change (S$_\mathrm{\Delta B_h}$), positive and negative change in $\mathrm{B_h}$, mid-time of the change in the field (T$_m$) since the start time of the flare, duration of the change in the field ($\tau$), start time of the change in the field (T$_s$) since the start time of the flare, and the rate of change in the field ($\mathrm\Delta B_h/\tau$). The area is expressed in millionths of the solar hemisphere, which is equivalent to 3$\times$10$^6$ km$^2$.} 
    \begin{tabular}{lccccccccc}
        \hline
    \hline
Event &	Flare start	time & Flare & $\mathrm{S_{\Delta B_h}}$  &  $+\Delta B_h$  & $-\Delta B_h$ & $\mathrm{T_m}$  & $\tau$ & $\mathrm{T_s}$  & $|\Delta B_h|/\tau$ \\
no. &       (UT)  & Class &  (MSH) &                 (G)  &          (G)         &  (min.)           & (min.)     & (min.) &       (G/min.)  \\
\hline
1 & 2010-08-07T17:55 & M1.0 & 82.8 & 141.0 $\pm$ 74.5 & 159.3 $\pm$ 92.7 & 24.4 & 12.9 & 14.7 & 13.5 \\
2 & 2011-02-15T01:43 & X2.2 & 77.3 & 259.6 $\pm$ 143.1 & 155.8 $\pm$ 104.2 & 13.3 & 10.0 & 7.0 & 20.2 \\
3 & 2011-08-03T13:17 & M6.0 & 125.9 & 169.2 $\pm$ 119.9 & 162.5 $\pm$ 105.0 & 25.2 & 13.3 & 15.6 & 13.9 \\
4 & 2011-09-06T22:11 & X2.1 & 29.5 & 277.7 $\pm$ 172.8 & 166.5 $\pm$ 119.1 & 8.5 & 7.3 & 4.2 & 34.3 \\
5 & 2011-10-02T00:37 & M3.9 & 9.8 & 164.5 $\pm$ 108.9 & 147.7 $\pm$ 74.5 & 10.8 & 8.2 & 5.4 & 22.0 \\
6 & 2011-11-15T12:29 & M1.9 & 7.0 & 129.1 $\pm$ 114.9 & 138.7 $\pm$ 116.9 & 10.7 & 6.8 & 6.5 & 22.0 \\
7 & 2011-12-27T04:11 & C8.9 & 7.6 & 134.0 $\pm$ 84.3 & 158.1 $\pm$ 89.7 & 9.9 & 6.3 & 5.9 & 26.2 \\
8 & 2012-01-23T03:37 & M8.7 & 146.3 & 175.3 $\pm$ 118.2 & 180.7 $\pm$ 123.8 & 27.2 & 14.0 & 17.4 & 14.5 \\
9 & 2012-03-07T00:01 & X5.4 & 125.6 & 202.4 $\pm$ 148.9 & 165.9 $\pm$ 121.9 & 16.3 & 10.9 & 9.2 & 19.6 \\
10 & 2012-03-09T03:21 & M6.3 & 110.2 & 154.0 $\pm$ 120.1 & 168.2 $\pm$ 122.5 & 25.1 & 14.8 & 14.9 & 12.5 \\
11 & 2012-03-10T17:15 & M8.4 & 135.9 & 168.5 $\pm$ 118.7 & 157.7 $\pm$ 124.7 & 33.4 & 17.5 & 22.3 & 10.8 \\
12 & 2012-03-14T15:07 & M2.8 & 6.1 & 165.6 $\pm$ 109.1 & 177.1 $\pm$ 119.3 & 14.3 & 6.8 & 9.4 & 28.3 \\
13 & 2012-07-12T15:37 & X1.4 & 492.9 & 172.2 $\pm$ 122.6 & 181.6 $\pm$ 125.9 & 57.0 & 18.2 & 45.9 & 11.4 \\
14 & 2012-11-21T06:45 & M1.4 & 14.9 & 144.9 $\pm$ 110.0 & 143.2 $\pm$ 93.4 & 11.3 & 7.8 & 6.1 & 19.3 \\
15 & 2013-04-11T06:55 & M6.5 & 25.0 & 158.1 $\pm$ 115.6 & 150.9 $\pm$ 108.5 & 16.6 & 9.9 & 8.8 & 17.5 \\
16 & 2013-05-16T21:35 & M1.3 & 23.7 & 135.8 $\pm$ 82.7 & 140.9 $\pm$ 96.7 & 13.8 & 9.3 & 8.2 & 16.1 \\
17 & 2013-05-31T19:51 & M1.0 & 1.5 & 133.9 $\pm$ 65.7 & 152.1 $\pm$ 94.3 & 7.3 & 5.7 & 4.5 & 25.1 \\
18 & 2013-08-17T18:49 & M1.4 & 149.6 & 165.8 $\pm$ 124.3 & 206.0 $\pm$ 136.3 & 35.4 & 18.7 & 24.4 & 12.1 \\
19 & 2013-12-28T17:53 & C9.3 & 3.0 & 137.5 $\pm$ 104.5 & 193.5 $\pm$ 82.2 & 7.7 & 5.2 & 5.0 & 32.0 \\
20 & 2014-01-07T18:03 & X1.2 & 149.6 & 142.4 $\pm$ 86.0 & 146.0 $\pm$ 96.6 & 25.4 & 14.0 & 15.9 & 12.2 \\
21 & 2014-01-31T15:31 & M1.1 & 15.3 & 140.7 $\pm$ 76.7 & 142.2 $\pm$ 92.6 & 10.2 & 6.7 & 5.8 & 21.5 \\
22 & 2014-02-01T07:13 & M3.0 & 34.8 & 147.5 $\pm$ 114.0 & 139.5 $\pm$ 88.8 & 10.0 & 7.0 & 5.5 & 22.9 \\
23 & 2014-02-12T03:51 & M3.7 & 62.3 & 166.5 $\pm$ 129.1 & 159.2 $\pm$ 123.1 & 22.7 & 11.3 & 14.3 & 16.4 \\
24 & 2014-03-20T03:41 & M1.7 & 20.8 & 125.3 $\pm$ 79.3 & 137.1 $\pm$ 95.9 & 11.9 & 7.8 & 6.2 & 18.6 \\
25 & 2014-08-01T17:55 & M1.5 & 71.1 & 166.6 $\pm$ 112.5 & 168.0 $\pm$ 118.2 & 23.1 & 12.1 & 14.3 & 15.7 \\
26 & 2014-08-25T14:45 & M2.0 & 76.6 & 144.8 $\pm$ 129.1 & 142.6 $\pm$ 85.2 & 21.1 & 12.7 & 12.2 & 12.7 \\
27 & 2014-08-25T20:05 & M3.9 & 33.7 & 189.1 $\pm$ 108.4 & 144.8 $\pm$ 63.8 & 10.7 & 7.4 & 5.7 & 23.4 \\
28 & 2014-09-08T23:11 & M4.5 & 294.7 & 156.4 $\pm$ 103.1 & 163.7 $\pm$ 108.4 & 53.3 & 18.5 & 42.8 & 9.8 \\
29 & 2014-09-10T17:21 & X1.6 & 130.8 & 175.7 $\pm$ 92.0 & 145.8 $\pm$ 91.2 & 25.6 & 13.2 & 15.6 & 13.2 \\
30 & 2014-09-28T02:39 & M5.1 & 86.0 & 164.8 $\pm$ 121.2 & 167.2 $\pm$ 118.8 & 18.2 & 11.2 & 10.5 & 17.0 \\
31 & 2014-10-22T14:01 & X1.6 & 300.2 & 151.6 $\pm$ 120.5 & 152.6 $\pm$ 113.4 & 24.0 & 12.6 & 15.2 & 13.5 \\
32 & 2014-12-17T00:57 & M1.5 & 32.1 & 151.3 $\pm$ 109.9 & 151.6 $\pm$ 108.1 & 11.0 & 7.7 & 6.3 & 21.4 \\
33 & 2014-12-17T04:25 & M8.7 & 161.2 & 167.7 $\pm$ 114.7 & 165.0 $\pm$ 130.8 & 25.7 & 13.3 & 14.9 & 14.0 \\
34 & 2014-12-18T21:41 & M6.9 & 121.0 & 171.6 $\pm$ 92.7 & 145.1 $\pm$ 105.8 & 22.3 & 12.6 & 14.0 & 13.9 \\
35 & 2014-12-20T00:11 & X1.8 & 246.6 & 185.7 $\pm$ 113.4 & 164.1 $\pm$ 109.2 & 18.6 & 11.9 & 10.8 & 15.8 \\
36 & 2015-11-04T13:31 & M3.7 & 38.5 & 140.2 $\pm$ 77.1 & 148.7 $\pm$ 72.2 & 19.7 & 10.2 & 12.8 & 15.6 \\
37 & 2015-11-09T12:49 & M3.9 & 47.4 & 160.7 $\pm$ 87.0 & 159.7 $\pm$ 103.3 & 17.5 & 10.0 & 10.2 & 16.9 \\
\hline
    \end{tabular}
    \label{tab:bh_change}
\end{table}

\begin{figure*}
    \centering
    \includegraphics[width=1\linewidth]{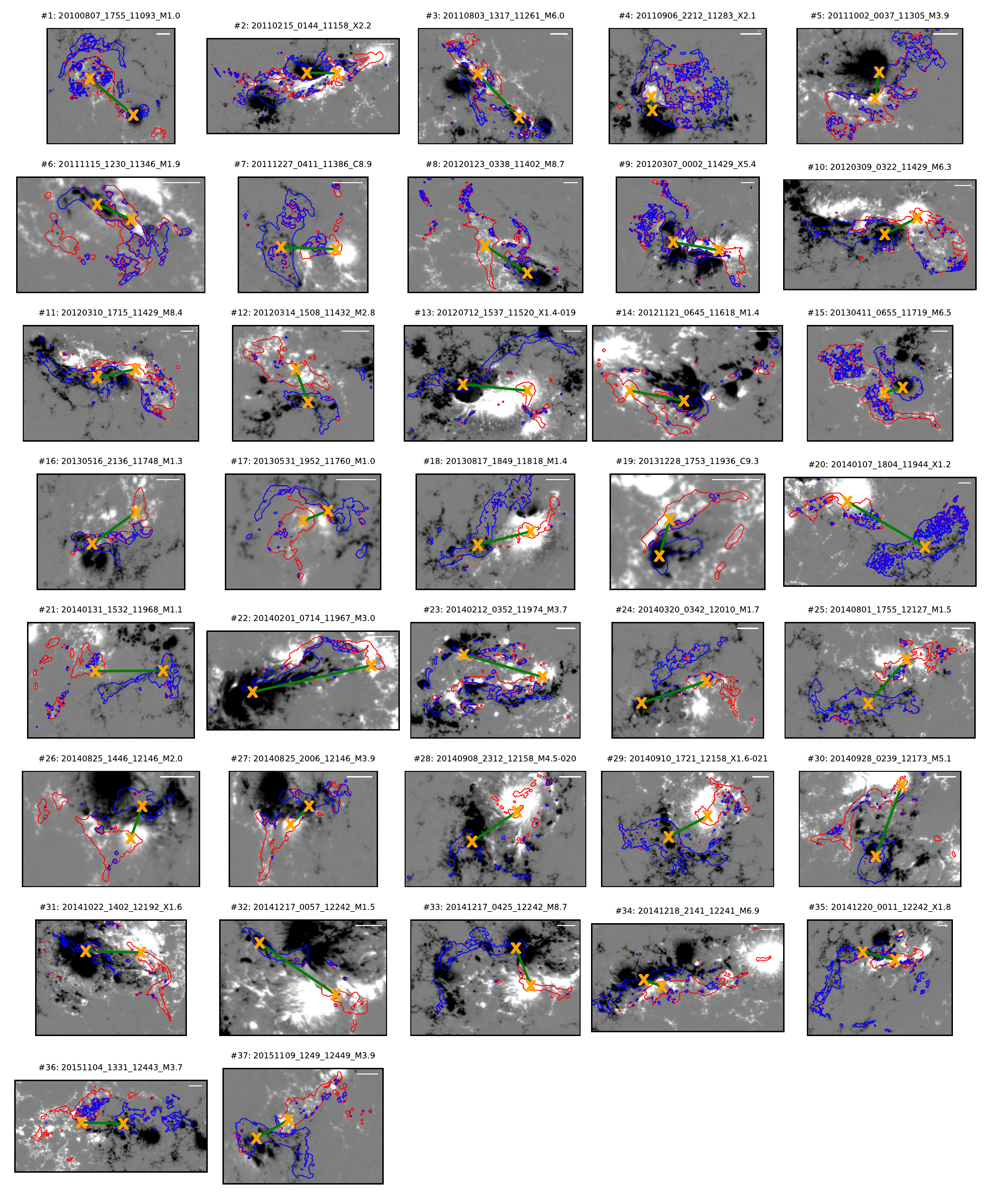}
 \caption{Overview of cumulative ribbons in 37 analyzed flares. The background image shows the $B_z$ component of the magnetic field, where black and white indicate negative and positive polarities (saturated at $\pm$800~G), respectively. Red and blue contours refer to the cumulative ribbons over the positive and negative polarities of B$_z$, respectively. The magnetic-flux weighted centroids of the ribbons are indicated by cross signs, which are connected by the solid green line. The solid horizontal white line in each panel indicates the length of 20\arcsec. The flare index is highlighted in the title of each panel.}
    \label{fig:appendix_overview_ribbon_sep}
\end{figure*}

\begin{figure*}
    \centering
    \includegraphics[width=1\linewidth]{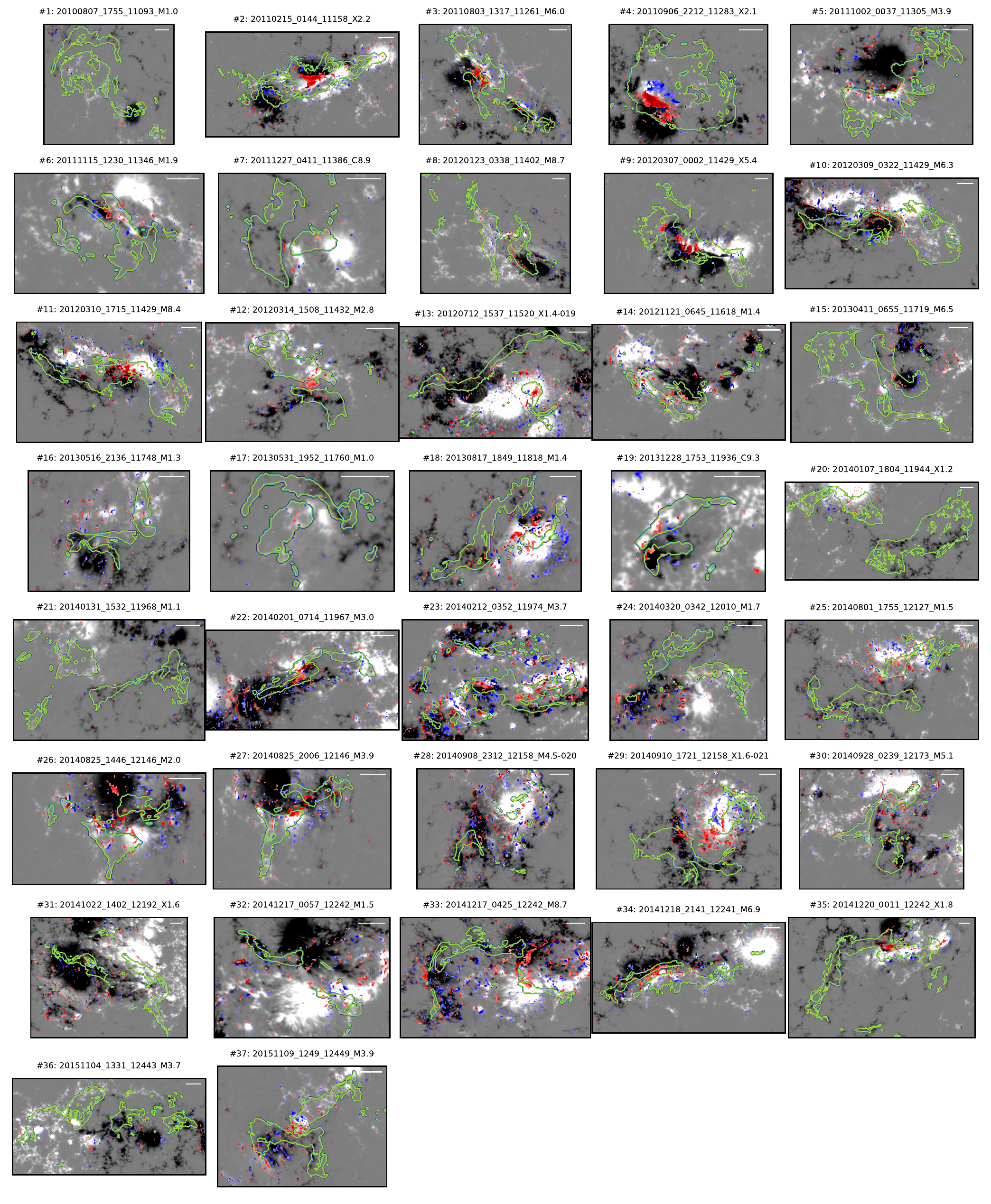}
    \caption{Overview of horizontal magnetic field changes in 37 analyzed flares. Each panel shows the difference in $B_h$ at flare start and end times, $\mathrm{\Delta B_h = B_h(T_e) - B_h(T_s)}$, where blue and red colors represent negative and positive changes (saturated at $\pm600$~G). The background image shows the $B_z$ component of the magnetic field, where black and white indicate negative and positive polarities (saturated at $\pm$800~G), respectively. The green contours refer to the cumulative flare ribbons. The solid horizontal white line in each panel indicates the length of 20\arcsec.}
    \label{fig:appendix_overview_bh_change}
\end{figure*}

\end{document}